\documentclass[12pt]{article}
\usepackage{epsfig}
\usepackage{amssymb}
\usepackage{axodraw}

\def\lsim{\mathrel{\rlap{\lower4pt\hbox{\hskip1pt$\sim$}}
    \raise1pt\hbox{$<$}}}                
\def\gsim{\mathrel{\rlap{\lower4pt\hbox{\hskip1pt$\sim$}}
    \raise1pt\hbox{$>$}}}                
\newcommand{\half}{$\frac{1}{2}$}        

\newcommand{\alphas}{\alpha_{\mathrm{s}}}
\newcommand{\alphaem}{\alpha_{\mathrm{em}}}
\newcommand{\pT}{p_{\perp}}
\newcommand{\ECM}{E_{\mathrm{CM}}}
\newcommand{\EPCM}{E'_{\mathrm{CM}}}
\newcommand{\NME}{N_{\mathrm{ME}}}
\newcommand{\NPS}{N_{\mathrm{PS}}}
\newcommand{\NPPS}{N'_{\mathrm{PS}}}

\renewcommand{\b}{\mathrm{b}}
\renewcommand{\c}{\mathrm{c}}
\renewcommand{\d}{\mathrm{d}}
\newcommand{\e}{\mathrm{e}}
\newcommand{\ee}{\ensuremath{\mathrm{e^+}\mathrm{e^-}}}
\newcommand{\f}{\mathrm{f}}
\newcommand{\g}{\mathrm{g}}
\newcommand{\hrm}{\mathrm{h}}
\newcommand{\p}{\mathrm{p}}
\newcommand{\q}{\mathrm{q}}
\newcommand{\s}{\mathrm{s}}
\renewcommand{\t}{\mathrm{t}}
\renewcommand{\u}{\mathrm{u}}
\newcommand{\A}{\mathrm{A}}
\renewcommand{\B}{\mathrm{B}}

\renewcommand{\H}{\mathrm{H}}

\newcommand{\W}{\mathrm{W}}
\newcommand{\Z}{\mathrm{Z}}
\newcommand{\bbar}{\overline{\mathrm{b}}}
\newcommand{\cbar}{\overline{\mathrm{c}}}
\newcommand{\dbar}{\overline{\mathrm{d}}}
\newcommand{\fbar}{\overline{\mathrm{f}}}

\newcommand{\qbar}{\overline{\mathrm{q}}}
\newcommand{\sbar}{\overline{\mathrm{s}}}
\newcommand{\tbar}{\overline{\mathrm{t}}}
\newcommand{\ubar}{\overline{\mathrm{u}}}

\newcommand{\sg}{\tilde{\mathrm{g}}}
\newcommand{\sq}{\tilde{\mathrm{q}}}
\newcommand{\sqbar}{\overline{\tilde{\mathrm{q}}}}
\newcommand{\st}{\tilde{\mathrm{t}}}
\newcommand{\stbar}{\overline{\tilde{\mathrm{t}}}}

\newcommand{\Py}{{\sc{Pythia}}}

\newenvironment{Itemize}{\begin{list}{$\bullet$}%
{\setlength{\topsep}{0.2mm}\setlength{\partopsep}{0.2mm}%
\setlength{\itemsep}{0.2mm}\setlength{\parsep}{0.2mm}}}%
{\end{list}}
\newcounter{enumct}

\newlength{\abstwidth}
\newlength{\captivewidth}
\begin{document}
 
\sloppy

\pagestyle{empty}

\begin{flushright}
LU TP 00--42\\
hep-ph/0010012\\
October 2000
\end{flushright}

\vspace{\fill}

\begin{center}
{\LARGE\bf QCD Radiation off Heavy Particles}\\[10mm]
{\Large E. Norrbin\footnote{emanuel@thep.lu.se} and %
T. Sj\"ostrand\footnote{torbjorn@thep.lu.se}} \\[3mm]
{\it Department of Theoretical Physics,}\\[1mm]
{\it Lund University,}\\[1mm]
{\it S\"olvegatan 14A,}\\[1mm]
{\it S-223 62 Lund, Sweden}
\end{center}
 
\vspace{\fill}
 
\begin{center}
{\bf Abstract}\\[2ex]
\begin{minipage}{\abstwidth}
We study QCD radiation in decay processes involving heavy particles. 
As input, the first-order gluon emission rate is calculated in a
number of reactions, and comparisons of the energy flow patterns show 
a non-negligible process dependence. To proceed further, the QCD parton 
shower language offers a convenient approach to include multi-gluon
emission effects, and to describe exclusive event properties.
An existing shower algorithm is extended to take into account the 
process-dependent mass, spin and parity effects, as given by the matrix 
element calculations. This allows an improved description of multiple 
gluon emission effects off b and t quarks, and also off nonstandard 
particles like squarks and gluinos. Phenomenological applications are 
presented for bottom  production at LEP, Higgs particle decay to heavy 
flavours, top production and decay at linear colliders, and some 
simple supersymmetric processes.  
\end{minipage}
\end{center}
 
\vspace{\fill}

\clearpage
\pagestyle{plain}
\setcounter{page}{1} 

\section{Introduction}

Heavy particles often tend to be in the focus of particle physics 
research. The search for new physics at the high-energy frontier
is an obvious example. Not only are the hypothetical new heavy 
states here of interest, but often their detection relies on decay 
chains that involve heavy flavours. Here the $\b$ quark is the prime 
example. By secondary vertices or semileptonic decays it can 
be tagged. It is a main decay product of top, where the $\b$ tag is
a standard requirement for top identification above background at 
hadron colliders. A light Standard Model Higgs 
($m_{\hrm} \lsim 140$~GeV) predominantly decays to $\b\bbar$, and
for any Higgs state the $\b\bbar$ branching ratio is a key parameter
in pinning down its nature. These tagging and identification aspects
also appear in other scenarios beyond the standard model.

Many of the potentially new particles carry colour, like squarks,
gluinos and leptoquarks. Also colourless states often decay to 
quarks. Therefore QCD radiation is unavoidable, and the large value 
of the $\alphas$ coupling implies that such radiation can be quite 
profuse. It is well-known, e.g. from LEP, that fixed-order 
perturbation theory fails to describe QCD radiation effects off light 
quarks: the rate of a few well separated jets can be reproduced,
but the internal structure of these jets requires the resummation
of multiple-gluon effects. The most successful method for achieving
such a resummation is to apply the parton-shower language, where
explicit final states can be generated, with full respect for 
energy--momentum conservation and other constraints. By introducing 
some low fixed
cut-off scale $Q_0 \approx 1$~GeV, the very soft and nonperturbative
r\'egime of QCD can be factored off and put in a universal hadronization 
description, such as the Lund string fragmentation model \cite{Lundstring}. 
With this approach, it is possible to obtain a quite accurate description 
of essentially all hadronic final-state properties of $\e^+\e^-$ 
annihilation events. The physics is more complicated in a hadron collider
environment, with additional effects e.g. from initial-state QCD radiation
and underlying events, so the level of ambition may have to be set 
accordingly. However, this applies to all conceivable descriptions, 
so again we expect a realistic description of jets to be most easily 
achieved with the help of the parton-shower language.

Heavy quarks radiate less than light ones in the soft-gluon region, so in 
some respects QCD multiple-emission corrections are not as crucial. 
The $\b$ is sufficiently light that it still radiates profusely, but top 
and supersymmetric particles may be heavy enough that multiple-gluon
emission effects are limited. Furthermore, for particles that are very 
short-lived, the width will provide a cut-off on soft-gluon emission
\cite{widtheffect}. So for new-particle searches, higher-order QCD effects 
may be less of an issue. However, whenever higher precision is required, 
like for a mass determination of a new state, it may still be
necessary to model whatever soft-gluon emission effects there are.
And, of course, the decays of these new states will bring further 
radiation, e.g. like $\t \to \b\W^+$, where the radiation off the $\t$ 
and off the $\b$ become intertwined, and where width effects may become 
important for short-lived particles. The aim of the current article 
is precisely to improve the description of gluon radiation off $\b$ 
and $\c$ quarks as well as off heavier objects, in order to allow 
higher-precision physics studies.

The starting point is the calculation of a large number of first-order
matrix elements, for gluon emission in decay processes,
often generalizing on results already found in the literature. The 
processes thereby covered include gluon emission e.g. in 
$\gamma^*/\Z^0/\hrm^0/\A^0 \to \b\bbar$ (with differences between a vector, 
axial vector, scalar and pseudoscalar source of the $\b$'s), 
$\t \to \b\W^+$, and various supersymmetric reactions. Quite apart from 
the implementation in the shower framework, these calculations can also
be used to assess the degree of (non)universality, i.e. the dependence
on the colour and spin structure of the processes under identical
kinematical conditions. In particular, we will show that the `dead cone' 
picture can be quite misleading for the emission of more energetic gluons.  

These matrix elements are taken as input to an improved version of
an existing shower algorithm \cite{Mats}, which primarily has been used 
for radiation off lighter quarks and gluons. This shower algorithm tends 
to slightly overestimate the amount of gluon radiation for 
$\e^+\e^- \to \gamma^*/\Z^0 \to \q\qbar\g$ when $\q$ is massless, so a 
simple rejection procedure can be used to match the hardest emission to 
the massless matrix elements. The radiation off heavier 
quarks is underestimated in the collinear region, however. Or, put
differently, the dead cone \cite{deadcone} effects are exaggerated. 
In order to allow a corresponding approach for heavy flavours, the 
evolution variable therefore has been modified to bring up the
emission rate. It therefore becomes possible to use the process-specific 
matrix elements as rejection factors also here, and furthermore to do it
in all stages of the shower.

The one place where precision tests are possible is in studies of
$\b$ production at LEP1 \cite{QCDWG}. We therefore study this topic in 
detail here, and use it as a test bed for different possible variations of 
the basic shower algorithm. Other studies include Higgs, top and 
supersymmetric particle production and decay. Furthermore, we also include 
some plots to illustrate differences between gluon radiation off sources 
of different spin. 

While not the main thrust of this article, we note that new data on
gluon branching to heavy flavours, $\g \to \c\cbar$ and  $\g \to \b\bbar$, 
has been presented by the LEP collaborations \cite{QCDWG}. We therefore also 
discuss what implications this might have for the shower algorithm, and
whether the data could be accommodated by reasonable modifications.
   
The plan of the article is as follows. In Section 2, the existing older
shower algorithm is explained, together with a few intermediate variants 
thereof. The new approach is described in Section 3. A survey of the 
matrix-element calculations and the consequent radiation patterns
is given in Section 4. Some applications are then presented in Section 5. 
The separate topic of gluon splitting is discussed in Section 6. Finally, 
a summary and outlook is given in Section 7.

\section{Previous Models}

Several shower algorithms have been proposed in the literature.
Today the three most commonly used ones probably are those found in  
\textsc{Pythia/Jetset} \cite{Mats,Pythia}, \textsc{Herwig} 
\cite{Herwig, Herwighvy} and \textsc{Ariadne} \cite{Ariadne}. The studies 
in this article will be based on the former one.

The \textsc{Pythia} final-state shower consists of an evolution in the 
squared mass $m^2$ of a parton. That is, emissions are ordered in decreasing 
mass of the radiating parton, and the Sudakov form factor \cite{Sudakov} is defined as the 
no-emission rate in the relevant mass range. Such a choice is not as 
sophisticated as the angular one in \textsc{Herwig} or the transverse 
momentum one in \textsc{Ariadne}, but usually the three tend to give 
similar results for $\e^+\e^-$ annihilation events. An exception, where 
small but significant differences were found, is the emission of photons 
in the shower \cite{photonemission}. In general, comparisons between the 
three are helpful in estimating a range of theoretical uncertainty,
in interpretations of existing data or in predictions for the future.

One of the advantages of the \textsc{Pythia} algorithm is that a mapping 
between the parton-shower and matrix-element variables is rather 
straightforward to $\mathcal{O}(\alphas)$ for massless quarks, and that 
already the basic shower populates the full phase space region very 
closely the same way as the matrix element. It is therefore possible to 
introduce a simple correction to the shower to bring the two into agreement. 
Also in \textsc{Ariadne} the massless matrix-element matching is 
straightforward. By contrast, the \textsc{Herwig} angular-ordered approach
does not automatically cover the full $\q\qbar\g$ phase space, which means 
that a subset of three-jet events are not generated at all. This leads to 
problems in the description of LEP data, that were overcome by separately 
adding the missing class of three-jet events \cite{Mikecorr}, in addition 
to matching to the matrix elements in the allowed region. More recently, 
a similar approach has been applied to top decay \cite{Miketdec}, where 
again \textsc{Herwig} did not populate the full phase space.

\subsection{The massless shower}

In addition to mass, the other main variable in the \textsc{Pythia} 
shower is $z$, as used in the splitting kernels. It is defined as the
energy fraction taken by the first daughter in the CM frame of the event.
That is, in a branching $a \to b+c$, $E_b = z E_a$ and $E_c = (1-z) E_a$.
In the original choice of $z$, which is done at the same time as
$m_a$ is selected, the $b$ and $c$ masses are not yet known, and are 
therefore imagined massless, also in cases where either of them is
known to have a non-vanishing on-shell mass. A cut-off scale 
$m_{\mathrm{min}} = Q_0 \approx 1$~GeV is used to constrain the allowed 
phase space, so that only branchings with $m_a > m_{\mathrm{min}}$ are 
allowed. For a massive quark, the cut-off is shifted to
\begin{equation} 
m_{a,\mathrm{min}} = \sqrt{m_a^2 + \frac{Q_0^2}{4}} + \frac{Q_0}{2}~.
\end{equation}
The allowed $z$ range, $z_- < z < z_+$, then becomes
\begin{equation} 
z_{\pm} = \frac{1}{2} \left\{ 1 \pm \beta_a \Theta(m_a - m_{a,\mathrm{min}})
\right\} ~,
\end{equation}
with $\beta_a = |\mathbf{p}_a|/E_a$ the $a$ velocity and $\Theta(x)$ the
step function.

At a later stage of the evolution, when $m_b$ and $m_c$ are being selected, 
possibly well above $Q_0$, the previously found $z$ may be incompatible with 
these masses. The adopted solution is to take into account mass effects by 
reducing the magnitude of the three-momenta $\mathbf{p}_b = - \mathbf{p}_c$ 
in the rest frame of $a$. Expressed in terms of four-momenta in an arbitrary 
frame, this is equivalent to
\begin{eqnarray}
p_b & = & (1 - k_b) p_b^{(0)} + k_c p_c^{(0)} ~, \nonumber\\
p_c & = & (1 - k_c) p_c^{(0)} + k_b p_b^{(0)} ~,
\label{shuffle}
\end{eqnarray}
where $p_b^{(0)}$ and $p_c^{(0)}$ are the original massless momenta 
and $p_b$ and $p_c$ the modified massive ones. The parameters $k_b$ 
and $k_c$ are found from the constraints $p_b^2 = m_b^2$ and 
$p_c^2 = m_c^2$:
\begin{equation} 
k_{b,c} = \frac{ m_a^2 - \lambda_{abc} \pm (m_c^2 - m_b^2)}{2 m_a^2} ~,
\end{equation}
with
\begin{equation}
\lambda_{abc} = \sqrt{ (m_a^2 - m_b^2 - m_c^2)^2 - 4 m_b^2 m_c^2 } ~. 
\end{equation}
The relation between the preliminary and final energy sharing thus is 
given by
\begin{equation}
z' = \frac{E_b}{E_a} = (1-k_b) z + k_c (1-z) = 
\frac{ m_a^2 - \lambda_{abc} + m_b^2 - m_c^2}{2 m_a^2} 
+ \frac{\lambda_{abc}}{m_a^2} z ~, 
\end{equation}
with $z = E_b^{(0)}/E_a$ as above. The transverse momentum $\pT$ 
of $b$ and $c$ with respect to the $a$ direction is given by 
\begin{equation} 
\pT^2 = \frac{\lambda_{abc}^2}{m_a^2} \left\{ \frac{z(1-z)}{\beta_a^2}
- \frac{1 - \beta_a^2}{4 \beta_a^2} \right\} ~,
\label{pTbranching}
\end{equation}
but is approximated by $\pT^2 \approx z(1-z) m_a^2$ when used as argument in
$\alphas(\pT^2)$ for the shower.

Angular ordering is not automatic, but is implemented by vetoing 
emissions that don't correspond to decreasing opening angles. The 
opening angle of a branching $a \to b+c$ is calculated approximately 
as
\begin{equation}
 \theta \approx \frac{p_{\perp b}}{E_b} + \frac{p_{\perp c}}{E_c}
 \approx \sqrt{z(1-z)} m_a \left( \frac{1}{z E_a} + 
 \frac{1}{(1-z) E_a} \right) 
 = \frac{1}{\sqrt{z(1-z)}} \frac{m_a}{E_a} ~.
\label{angle}
\end{equation}

\begin{figure}[t]
\begin{center}
\begin{picture}(220,130)(0,-65)
\ArrowLine(30,-20)(60,0)\Text(18,-20)[]{$\e^+$}
\ArrowLine(60,0)(30,20)\Text(18,20)[]{$\e^-$}
\Photon(60,0)(100,0){3}{3}\Text(80,12)[]{$\Z^0$}
\Text(115,-20)[]{1}\Text(145,-40)[]{3}
\ArrowLine(200,-60)(100,0)\Text(210,-60)[]{7}
\Text(115,20)[]{2}\Text(145,40)[]{5}
\ArrowLine(100,0)(200,60)\Text(210,60)[]{11}
\Gluon(160,-36)(200,-45){3}{4}\Text(210,-45)[]{8}
\Gluon(130,-18)(160,-20){3}{3}\Text(145,-9)[]{4}
\ArrowLine(200,-30)(160,-20)\Text(210,-30)[]{9}
\ArrowLine(160,-20)(200,-10)\Text(210,-10)[]{10}
\Gluon(130,18)(160,20){3}{3}\Text(145,9)[]{6}
\Gluon(160,20)(200,10){3}{4}\Text(210,10)[]{14}
\Gluon(160,20)(200,30){3}{4}\Text(210,30)[]{13}
\Gluon(160,36)(200,45){3}{4}\Text(210,45)[]{12}
\Text(100,-60)[]{(a)}
\end{picture}
\hspace{10mm}
\begin{picture}(170,130)(50,-65)
\Photon(60,0)(100,0){3}{3}\Text(80,12)[]{0}
\ArrowLine(180,-50)(100,0)\Text(200,-50)[]{1 ($\q$)}
\Text(120,-25)[]{$i$}
\ArrowLine(100,0)(180,50)\Text(200,50)[]{2 ($\qbar$)}
\Gluon(140,-25)(180,-10){3}{4}\Text(200,-10)[]{3 ($\g$)}
\Text(125,-60)[]{(b)}
\end{picture}
\caption{Example of showers, with the notations used in the text.
(a) A generic shower. (b) A shower giving a three-jet event.}
\label{fig:Zshower}
\end{center}
\end{figure}
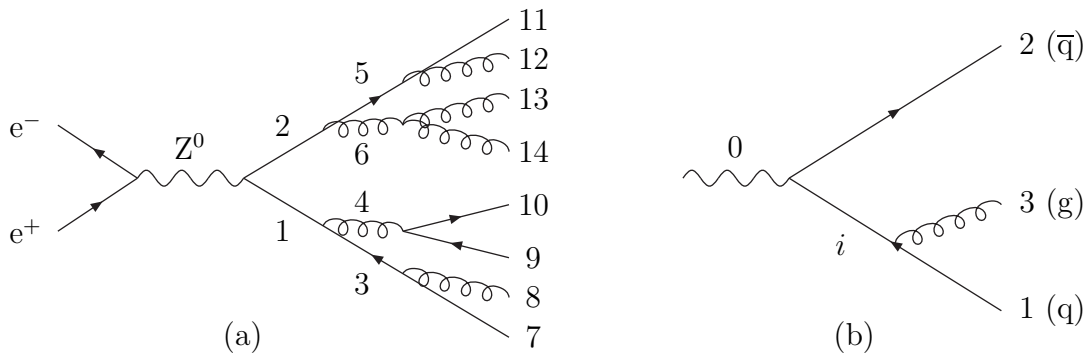

The procedure thus is the following, Fig.~\ref{fig:Zshower}a. 
In the $\gamma^*/\Z^0$ decay,
the two original partons 1 and 2 are produced, back-to-back in
the rest frame of the pair. In a first step, they are evolved 
downwards from a maximal mass equal to the CM energy, with the 
restriction that the two masses together should be below this CM 
energy. When the two branchings are found, they define $m_1$ and 
$m_2$ and the $z$ values of $1 \to 3+4$ and $2 \to 5+6$. These 
branchings obviously have smaller opening angles than the 
$180^{\circ}$ one between 1 and 2, so no angular-ordering constraints
appear here. A matching procedure to the matrix element is used
to correct the branchings, however, as will be described below.
In subsequent steps, a pair of partons like 3 and 4 are evolved in
parallel, from maximum masses given by the smaller of the mother (1)
mass and the respective daughter (3 or 4) energy. Here angular ordering 
restricts the region of allowed $z$ values in their branchings,
but there are no matrix-element corrections. Once $m_3$ and $m_4$ are 
fixed, the kinematics of the $1 \to 3+4$ branching needs to be 
modified according to eq.~(\ref{shuffle}). This is the reason
why the evolution is always done for a pair of partons (whereof 
not both need branch further, however), and why the final kinematics
of a branching is postponed to a later stage than the choice of $z$
value.  

Several other aspects of the shower could be discussed, such as the 
choice of non-isotropic azimuthal angles to improve the coherence
description and include gluon spin effects, the possibility also to 
emit photons, or the option to force some branchings in order to 
match higher-order matrix elements \cite{Johan}. These are of less 
interest here and thus not covered.

\subsection{The massless matrix element correction}

Let us now compare the parton-shower (PS) population of three-jet 
phase space with the matrix-element (ME) one, for the case of 
$\e^+\e^-$ annihilation to massless quarks. With the conventional 
ME numbering  $\q(1) \, \qbar(2) \, \g(3)$, with $x_j = 2 E_j / \ECM$, 
the matrix element is of the form \cite{MEqqgmassless}
\begin{equation}
\frac{1}{\sigma_0} \, \frac{\d \sigma_{\mathrm{ME}}}%
{\d x_1 \, \d x_2} = \frac{\alphas}{2\pi} \, 
C_F \, \frac{x_1^2 + x_2^2}{(1-x_1)(1-x_2)} ~,
\label{MEmassless}
\end{equation}
with the colour factor $C_F = 4/3$. We have normalized to the 
lowest-order cross section $\sigma_0$, so that the expression can 
be interpreted as a probability distribution. For future reference,
we will use the notation 
\begin{equation}
\frac{1}{\sigma_0} \frac{\d \sigma}%
{\d x_1 \, \d x_2} = \frac{\alphas}{2\pi} \, 
C_F \, \frac{N(x_1,x_2,r)}{(1-x_1)(1-x_2)} ~,
\label{genform}
\end{equation}
where $r = m_{\q}/\ECM = m_{\qbar}/\ECM$.
Thus $\NME(x_1, x_2, 0) = x_1^2 + x_2^2$.
 
There are two shower histories that could give such a three-jet
event. One is $\gamma^*/\Z^0(0) \to \q(i) \qbar(2) \to
\q(1) \qbar(2) \g(3)$, i.e. with an intermediate ($i$) quark 
branching $\q(i) \to \q(1) \g(3)$, illustrated in 
Fig.~\ref{fig:Zshower}b. 
This gives
\begin{eqnarray}
Q^2 & = & m_i^2 = (p_0 - p_2)^2 = (1-x_2) \ECM^2  ~,  
\\
z & = & \frac{p_0 p_1}{p_0 p_i} = \frac{E_1}{E_i} = 
\frac{x_1}{x_1 + x_3} = \frac{x_1}{2 - x_2}  ~, 
\\
& \Rightarrow & \frac{\d Q^2}{Q^2} \, \frac{\d z}{1-z} = 
\frac{\d x_2}{1-x_2} \, \frac{\d x_1}{x_3} ~.
\label{masslessJacobian}
\end{eqnarray}
The parton-shower probability for such a branching is
\begin{equation}
\frac{\alphas}{2\pi} \, C_F \, \frac{1+z^2}{1-z} \, 
\d z \, \frac{\d Q^2}{Q^2} =
\frac{\alpha_s}{2\pi} \, C_F \,
\left[ 1 + \left( \frac{x_1}{2-x_2} \right)^2 \right] 
\frac{1-x_1}{x_3} \frac{\d x_1 \, \d x_2}{(1-x_1)(1-x_2)} ~.
\label{NPSmassless}
\end{equation}
In the second shower history, the r\^oles of $\q$ and
$\qbar$ are interchanged, i.e. $x_1 \leftrightarrow x_2$.
This is the same set of Feynman graphs as in the matrix-element 
description, except that the shower does not include any interference 
between the two diagrams. The two shower expressions can therefore
be added to give the overall shower population of the three-jet phase 
space, of the form in eq.~(\ref{genform}) but with
\begin{equation}
\NPS(x_1,x_2,0) = \frac{1-x_1}{x_3} \left( 1 + 
\left( \frac{x_1}{2-x_2} \right)^2 \right) + \frac{1-x_2}{x_3} 
\left( 1 + \left( \frac{x_2}{2-x_1} \right)^2 \right) ~,
\end{equation}

\begin{figure}[t]
\begin{center}
\epsfig{file=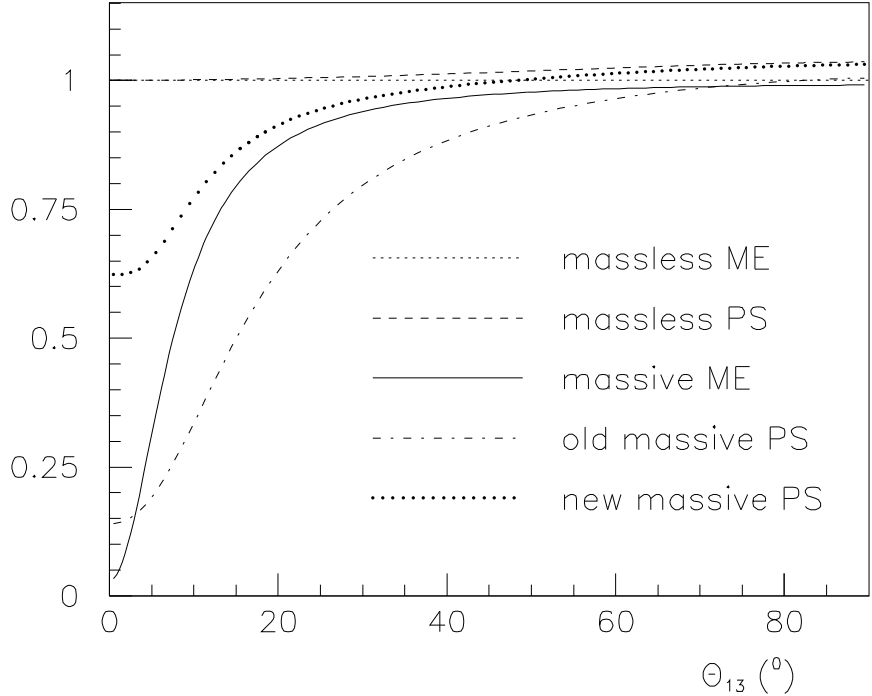}
\end{center}
\vspace{-2\baselineskip}
\caption{The gluon emission rate as a function of emission angle
$\theta_{13} = \theta_{\q\g}$, for a 10~GeV gluon energy at 
$\ECM = 91$~GeV, and with $m_b = 4.8$~GeV.
All curves are normalized to the massless 
matrix-element expression, eq.~(\protect\ref{MEmassless}),
here thus represented by the small-dotted line at unity.
Dashed: the massless shower before correction, 
$\NPS(x_1,x_2,0)/\NME(x_1,x_2,0) = 1/R_{\mathrm{ME/PS}}(x_1,x_2,0)$. 
Full: the rate from massive matrix elements,
$\NME(x_1,x_2,r)/\NME(x_1,x_2,0)$. 
Dash-dotted: the rate from massive parton shower,
$\NPS(x_1,x_2,r)/\NME(x_1,x_2,0)$. 
Large dots: the new rate from massive parton showers, 
$\NPPS(x_1,x_2,r)/\NME(x_1,x_2,0)$.}
\label{fig:suprmass}
\end{figure}
   
In spite of $\NPS$ being lengthier than $\NME$, it 
turns out that the two almost exactly agree over the whole phase 
space, but with the shower rate somewhat above, see e.g. 
Fig.~\ref{fig:suprmass}. It is therefore straightforward and efficient 
to use the ratio 
\begin{equation}
R_{\mathrm{ME/PS}}(x_1,x_2,0) =
\frac{\d \sigma_{\mathrm{ME}}}{\d \sigma_{\mathrm{PS}}}
= \frac{\NME(x_1,x_2,0)}{\NPS(x_1,x_2,0)}
\label{masslesscorr}
\end{equation}
as an acceptance factor inside the shower evolution, in order 
to correct the first emission of the quark and antiquark to
give a sum in agreement with the matrix element. 

Clearly, the shower will contain further branchings that 
modify the simple result, e.g. by the emission both from 
the $\q$ and the $\qbar$, but these effects are formally of 
$\mathcal{O}(\alphas^2)$ and thus beyond the accuracy 
we strive to match. One should also note that the shower modifies
the distribution in three-jet phase space by the appearance
of Sudakov form factors, and by using a running $\alphas (\pT^2)$ 
rather than a fixed one. In both these respects, however, the shower 
should be an improvement over the fixed-order result.

\subsection{The massive matrix element correction}
\label{sec:old}

The prescription of correcting the first branchings by the factor
in eq.~(\ref{masslesscorr}) was the original one, used up
until \textsc{Jetset}~7.3, for massless and massive quarks alike,
where the $x_i = 2E_i/\ECM$ variables were defined with masses 
included. In version 7.4 an intermediate `improvement' was
introduced, in that the massive matrix element expression was used,
as given for a vector source like the $\gamma^*$
\cite{MEqqgmassive}:
\begin{equation}
\NME( x_1, x_2, r)
= x_1^2 + x_2^2 - 4 r^2 x_3 - 8 r^4 - (2 r^2 + 4 r^4) 
\left( \frac{1-x_2}{1-x_1} + \frac{1-x_1}{1-x_2} \right) ~. 
\end{equation}
The shower algorithm itself was not changed, nor the assumed shower
weight, i.e. an acceptance factor 
$\NME(x_1,x_2,r)/\NPS(x_1,x_2,0)$ was
applied for the first branching on either side.
(The older behaviour remained as an option.)

The mass suppression in the matrix element is illustrated in
Fig.~\ref{fig:suprmass}. We remind that, in the soft-gluon 
limit, a spin-independent (and thus universal) eikonal expression
holds \cite{Herwighvy}:
\begin{eqnarray}
\frac{\d\sigma_{\q\qbar\g}}{\sigma_{\q\qbar}} & \propto & 
(-1) \left( \frac{p_1}{p_1 p_3} - \frac{p_2}{p_2 p_3} \right)^2 
\, \frac{\d^3 p_3}{E_3}
\nonumber \\
& = & \left( \frac{2 p_1 p_2}{(p_1 p_3) (p_2 p_3)} - 
\frac{m_1^2}{(p_1 p_3)^2} - \frac{m_2^2}{(p_2 p_3)^2} \right) 
\, E_3 \d E_3 \, \d \cos\theta_{13} ~.
\label{eikonalexpr}
\end{eqnarray}
In the limit of small angles $\theta_{13}$ this gives a mass
suppression factor
\begin{equation}
\frac{\d\sigma(x_3,\theta_{13},r)}{\d\sigma(x_3,\theta_{13},0)} 
\approx \left( \frac{\theta_{13}^2}{\theta_{13}^2 + 4 r^2} 
\right)^2 ~,
\label{deadconesuppr}
\end{equation}
i.e. the characteristic dead cone of opening angle approximately
$2r = m_{\q}/E_{\q}$. Note that the mapping between $(x_1 , x_2)$
and $(x_3, \theta_{13})$ depends on $r$, so 
eq.~(\ref{deadconesuppr}) is not quite the same as 
$\NME(x_1,x_2,r) / \NME(x_1,x_2,0)$.

\subsection{The massive phase space correction in the shower}
\label{sec:older}

More recently \cite{QCDWG}, the issue of masses
in the shower was further studied, since the expression for
$\NPS(x_1,x_2,0)$ had not been touched when 
$\NME$ was improved. 

In the derivation of $\NPS(x_1,x_2,r)$, one can start 
from the ansatz
\begin{eqnarray}
x_2 & = & 1 - \frac{m_i^2 - m_{\q}^2}{\ECM^2} ~,\nonumber
\\
x_1 & = & \left( 1 + \frac{m_i^2 - m_{\q}^2}{\ECM^2} \right)
 \left( (1-k_1) z + k_3(1-z) \right) ~, 
\label{shuffleqqg}
\\
x_3 & = & \left( 1 + \frac{m_i^2 - m_{\q}^2}{\ECM^2} \right)
 \left( (1-k_3) (1-z) + k_1 z \right) ~. \nonumber
\end{eqnarray}
The quark mass enters both in the energy sharing
between the intermediate quark $i$ and the antiquark 2,
and in the correction procedure of eq.~(\ref{shuffle})
for the splitting of energy in the branching $q(i) \to q(1) g(3)$.
The constraints $p_1^2 = m_{\q}^2$ and $p_3^2 =0$ give
$k_1 = 0$ and $k_3 = m_{\q}^2/m_i^2$. One then obtains
\begin{eqnarray}
Q^2 & = & m_i^2 = (1-x_2 + r^2) \ECM^2  ~,
\\
z & = & \frac{1}{2 - x_2} \left( x_1 - r^2 \, 
\frac{x_3}{1 - x_2} \right) ~,
\label{znewdef}
\\
& \Rightarrow & \frac{\d Q^2}{Q^2} \, \frac{\d z}{1-z} = 
\frac{\d x_2}{1-x_2 + r^2} \, \frac{\d x_1}{x_3} ~.
\end{eqnarray}
This gives the answer
\begin{eqnarray}
\NPS(x_1,x_2,r) & = & \frac{1-x_1}{x_3} 
\frac{1-x_2}{1 - x_2 + r^2} \left[ 1 + \frac{1}{(2-x_2)^2} 
\left( x_1 - r^2 \frac{x_3}{1-x_2} \right)^2 \right] 
\nonumber \\
& + & \left\{ x_1 \leftrightarrow x_2 \right\} ~,
\end{eqnarray}
where the second term comes from the graph where the antiquark 
radiates. 

The mass effects go in the `right' direction,
$\NPS(x_1,x_2,r) < \NPS(x_1,x_2,0)$, 
but actually so much so that
$\NPS(x_1,x_2,r) < \NME(x_1,x_2,r)$ in 
major regions of phase space. This is illustrated in 
Fig.~\ref{fig:suprmass}. Very crudely, one could say that the
massive shower exaggerates the angle of the dead cone by about 
a factor of two, in this rather typical example. There is no
dead cone as such built in, however, but a rather more coincidental
mass suppression mainly generated by the factor
$(1-x_2)/(1 - x_2 + r^2)$. 

Thus the amount of gluon emission off massive quarks is 
underestimated already in the original prescription, where masses 
entered in the kinematics but not in the ME/PS correction factor. 
When the intermediate `correction' ratio 
$\NME(x_1,x_2,r)/\NPS(x_1,x_2,0)$ is applied, 
the net result is a distribution even more off from the correct
one, by a factor $\NPS(x_1,x_2,r)/\NPS(x_1,x_2,0)$. Thus it would
have been better not to introduce the mass correction in JETSET 7.4.

Based on the results above, one can now instead use the correct 
ME/PS factor $\NME(x_1,x_2,r)/\NPS(x_1,x_2,r)$. A technical problem 
is that this ratio can exceed unity, in the example of 
Fig.~\ref{fig:suprmass} by up to almost a factor of two. This could
be solved e.g. by enhancing the raw rate of emissions by this 
factor. However, another trick was applied, based on the facts that
the shift of eqs.~(\ref{shuffle}) and (\ref{shuffleqqg}) implies 
that smaller-energy gluons would be allowed for a massive quark than 
a massless one, and additionally that the accessible $z$ range is 
overestimated in the original ansatz. Therefore, without any 
(noticeable) loss of phase space, $z$ can be rescaled to a $z'$ 
according to
\begin{equation}
(1-z') = (1-z)^k ~, ~~~\mathrm{with}~~~ 
k = \frac{\ln(r^2)}{\ln(Q_0^2/\ECM^2)} < 1 ~.
\end{equation}
The ME/PS correction factor then has to be compensated by $k$,
and thereby comes below unity almost everywhere --- the remaining
weighting errors are too small to be relevant.

This procedure, default since \textsc{Pythia} version 6.130, improves 
the shower description of mass effects in the amount of
three-jet events \cite{QCDWG}. Mass effects are only included correctly 
for the first branching of the $\q$ and $\qbar$ in the shower, however.
Subsequent emissions involve no correction procedure. Instead the dead
cone effect is exaggerated, similarly to what was shown in 
Fig.~\ref{fig:suprmass}. Furthermore, even for the first branchings,
only the possibility of a vector source decaying to two identical-mass
quarks is included, while the $\Z^0$ actually is a mixture of vector and 
axial vector, and e.g. the $\W^{\pm}$ would decay to two unequal-mass 
quarks. We will therefore next try to develop a more powerful and general 
approach.

\section{A New Approach}
\label{sec:new}
One of the advantages of the Monte Carlo approach is that, so long as
some upper estimate can be found that allows simple generation, rejection 
down to a more complex expression is straightforward. In particular,
for an evolution in some variable $Q$ with a (Sudakov) form factor 
built up from the no-emission probability, the `veto algorithm' 
\cite{Pythia} can be 
used, in which the initial overestimate of the emission rate is compensated 
by the possibility of a rejection, with a continued evolution downwards in 
$Q^2$ from the rejected value onwards. As we have seen, the current set of
$Q^2$ and $z$ variables is not so convenient, since the emission rate off
massive quarks is an underestimate of the correct rate, and therefore
the standard procedure does not work except after some extra tricks.
We will therefore pick another set of variables, preferably such that
they reduce to the old ones in the massless limit. Several approaches
could have been taken, but here we have chosen the minimal one of
retaining the $z$ definition and thereby the existing kinematics machinery.
As we will show, a modification for the branching $a \to bc$ from
$Q^2 = m_a^2$ to $Q^2 = m_a^2 - (m_a^2)_{\mathrm{on-shell}}$ is enough 
to do the job.

\subsection{The choice of shower variables}

Again consider $\e^+\e^- \to \gamma^*/\Z^0 \to \q(1) \, \qbar(2) \, \g(3)$,
and the formulae of eq.~(\ref{shuffleqqg}), but now assume 
$Q^2 = m_i^2 - m_{\q}^2 = m_i^2 - m_1^2$ for the $\q(i) \to \q(1) \g(3)$ 
branching. That corresponds to $1/Q^2$ being the propagator of the 
off-shell parton 1. Then eq.~(\ref{znewdef}) is unchanged while
\begin{eqnarray}
\label{eq:newQ2}
Q^2 & = & m_i^2 - m_1^2 = (1-x_2) \ECM^2  ~,
\\
& \Rightarrow & \frac{\d Q^2}{Q^2} \, \frac{\d z}{1-z} = 
\frac{\d x_2}{1-x_2} \, \frac{\d x_1}{x_3} ~.
\end{eqnarray}
So, with this simple trick, we have recovered the Jacobian of 
the massless case, eq.~(\ref{masslessJacobian}). This means that 
there is little mass suppression left in the shower evolution proper:
\begin{eqnarray}
\NPPS(x_1,x_2,r) & = & \frac{1-x_1}{x_3} \left[ 1 + \frac{1}{(2-x_2)^2} 
\left( x_1 - r^2 \frac{x_3}{1-x_2} \right)^2 \right] 
\nonumber \\
& + & \left\{ x_1 \leftrightarrow x_2 \right\} ~,
\end{eqnarray}
as can be seen in Fig.~\ref{fig:suprmass}. In particular, note that
$\NPPS(x_1,x_2,r)$ reduces to $\NPS(x_1,x_2,0)$, eq.~(\ref{NPSmassless}),
in the soft limit $x_3 \to 0$. Therefore a matrix-element correction
now is even more required, but also simpler to implement, to bring down 
the rate to a reasonable level.

So far, we have assumed decay to two equal-mass quarks, like in
$\Z^0$ decay. For many of the examples to be discussed, like $\W^{\pm}$
or $\t$ decay, the two decay products have unequal masses. It is also 
convenient to allow slightly unequal masses for particles with 
non-negligible widths, e.g. in $\e^+\e^- \to \t\tbar$. We will therefore
generalize to the case with 
$r_1 = m_1/\ECM = m_{\q}/\ECM \neq r_2 = m_2/\ECM = m_{\qbar}/\ECM$.
The range of kinematically allowed $x_i$ values is then 
$2r_1 \leq x_1 \leq 1 + r_1^2 - r_2^2$, 
$2r_2 \leq x_2 \leq 1 + r_2^2 - r_1^2$,
$0 \leq x_3 \leq 1 - (r_1 + r_2)^2$, with the joint condition that
\begin{equation}
(2 - 2x_1 - 2x_2 + x_1 x_2 + 2r_1^2 + 2 r_2^2)^2 \leq
(x_1^2 - 4 r_1^2) (x_2^2 - 4 r_2^2)  ~.
\end{equation}
The ansatz of eq.~(\ref{shuffleqqg}) is modified by $m_{\q} \to m_2$
while $k_3 \to m_1^2/m_i^2$. Furthermore
\begin{eqnarray}
Q^2 & = & m_i^2 - m_1^2 = (p_0 - p_2)^2 - m_1^2 = 
(1 + r_2^2 - r_1^2 - x_2) \ECM^2  ~,
\label{Qdiffmass}
\\
z & = & \frac{1}{2 - x_2} \left( x_1 - r_1^2 \, 
\frac{x_3}{1 + r_2^2 - r_1^2 - x_2} \right) ~,
\label{zdiffmass}
\\
& \Rightarrow & \frac{\d Q^2}{Q^2} \, \frac{\d z}{1-z} = 
\frac{\d x_2}{1 + r_2^2 - r_1^2 - x_2} \, \frac{\d x_1}{x_3} ~.
\end{eqnarray}
When a colourless particle decays to two colour triplets, the two possible 
one-gluon-emission shower histories then add up to give
\begin{equation}
\frac{1}{\sigma_0} \, \frac{\d \sigma_{\mathrm{PS}}}{\d x_1 \, \d x_2}
= \frac{\alphas}{2\pi} \, C_F \left\{ 
\frac{1 + z_1^2}{x_3 (1 + r_2^2 - r_1^2 - x_2)} +
\frac{1 + z_2^2}{x_3 (1 + r_1^2 - r_2^2 - x_1)} \right\} ~.
\label{PSdiffmass}
\end{equation}
Here $z_1$ is given by eq.~(\ref{zdiffmass}) and $z_2$ is obtained by 
exchanging $1 \leftrightarrow 2$. Since the numerators 
$1 \leq 1 + z_i^2 \leq 2$, their exact form is not a main concern for the 
qualitative discussions. As it turns out, the Monte Carlo procedure is 
simplified if the shower is generated with a numerator 2 instead of 
$1 + z_i^2$, so we will make this replacement in the following, with
the matrix element correction procedure compensating this overestimate. 

\subsection{The matrix element correction}

Next, one should try to relate this to the structure of the matrix elements
under similar conditions. We may expect graphs where either of the partons 
can radiate a gluon and therefore can give a propagator 
$1/Q_1^2 = 1/(m_{13}^2 - m_1^2)$, cf. eq.~(\ref{Qdiffmass}),
or $1/Q_2^2 = 1/(m_{23}^2 - m_2^2)$, depending on which side radiates. 
(Diagrams with a four-boson vertex, such as $\gamma^*/\Z^* \to \st\stbar\g$, 
are not singular and therefore do not affect the discussion.) After summing 
and squaring, the cross section then should have the form
\begin{eqnarray}
\frac{1}{\sigma_0} \, \frac{\d \sigma_{\mathrm{ME}}}{\d x_1 \, \d x_2} 
& = & \frac{\alphas}{2\pi} \, C_F \left\{ 
\frac{\ECM^4 A(x_1,x_2)}{(m_{13}^2 - m_1^2)^2} +
\frac{\ECM^4 B(x_1,x_2)}{(m_{23}^2 - m_2^2)^2} +
\frac{\ECM^4 C(x_1,x_2)}{(m_{13}^2 - m_1^2)(m_{23}^2 - m_2^2)} \right\}
\nonumber \\
& = & \frac{\alphas}{2\pi} \, C_F \left\{
\frac{A(x_1,x_2)}{(1 + r_2^2 - r_1^2 - x_2)^2} +
\frac{B(x_1,x_2)}{(1 + r_1^2 - r_2^2 - x_1)^2} \right. \\
& & \left. + \frac{C(x_1,x_2)}{(1 + r_2^2 - r_1^2 - x_2)%
(1 + r_1^2 - r_2^2 - x_1)} \right\}  ~. \nonumber 
\end{eqnarray}
The individual functions $A$, $B$ and $C$ depend on the gauge choice,
but the total cross section of course is gauge-independent. It also has
some general features, which can be seen in the soft-gluon limit expression, 
eq.~(\ref{eikonalexpr}). Here the interference term is giving a positive 
expression and the seemingly quadratically-divergent terms are negative and 
drive the cross section to zero in the collinear limit, the dead cone. 
In order to have an upper estimate of the cross section, for Monte Carlo 
applications, it is thus enough to have the singularity structure of the 
interference term modelled, and no need to worry about quadratic divergences. 

As before, we could now compare the total shower and total matrix element
rates, and define a corrective factor between the two. To simplify some of
the continued studies, however, we have adopted an alternative approach. 
Instead of adding the two shower histories to compare with the full matrix
element, one can split  the matrix element in two, such that each part
can be compared with only one shower history. Such a subdivision of course
is arbitrary, but should still be sensible. A gluon emitted close to the
$p_1$ direction should predominantly be emitted by parton 1, and vice versa.
A suitable such subdivision is in the proportions given by the propagators,
$1/Q_1^2 : 1/Q_2^2$, which also is the proportions between the two shower 
histories (in the $1 + z^2 \to 2$ approximation). For massless partons, and 
going to the soft-gluon limit, this means a probability 
$(1 + \cos\theta_{13})/2 = (1 - \cos\theta_{23})/2$ 
for emission off parton 1 and the rest for emission off parton 2.  

In the description of the gluon emission rate off parton 1, a matrix element 
fraction
\begin{equation}
W_{\mathrm{ME,1}}(x_1,x_2) = \frac{Q_2^2}{Q_1^2 + Q_2^2} \, 
\frac{1}{\sigma_0} \, \frac{\d \sigma_{\mathrm{ME}}}{\d x_1 \, \d x_2}
= \frac{1 + r_1^2 - r_2^2 - x_1}{x_3} \, 
\frac{1}{\sigma_0} \, \frac{\d \sigma_{\mathrm{ME}}}{\d x_1 \, \d x_2}
\end{equation}
should be compared with the first half of the total parton shower expression 
in eq.~(\ref{PSdiffmass}),
\begin{equation}
W_{\mathrm{PS,1}}(x_1,x_2) = \frac{\alphas}{2\pi} \, C_F 
\frac{2}{x_3 (1 + r_2^2 - r_1^2 - x_2)}  ~.
\label{PSfirsthist}
\end{equation}
The ME/PS correction factor then becomes
\begin{eqnarray}
R_1(x_1,x_2) & = &
\frac{W_{\mathrm{ME,1}}(x_1,x_2)}{W_{\mathrm{PS,1}}(x_1,x_2)} 
\nonumber \\
& = & \frac{(1 + r_2^2 - r_1^2 - x_2)(1 + r_1^2 - r_2^2 - x_1)}{2} \,
\left( \frac{\alphas}{2\pi} \, C_F \right)^{-1} 
\frac{1}{\sigma_0} \, \frac{\d \sigma_{\mathrm{ME}}}{\d x_1 \, \d x_2} ~,
\label{massivecorr}
\end{eqnarray}
which reduces to $R_1(x_1,x_2) = \NME(x_1,x_2,r)/2$ for $r_1 = r_2 = r$. 

The intention is that $R_1(x_1,x_2)$ should be finite and well-behaved
over all of phase space, with one factor of each divergence now multiplied 
on to the matrix element expression. We will illustrate the behaviour later 
on, but the key observation is that, for all the matrix elements we will 
study, searches over the full mass parameter plane and phase space have 
failed to find any point where the ratio is above unity. $R_1$ therefore
serves well as simple rejection factor. (Without the replacement 
$1 + z^2 \to 2$ in the shower description one would find 
$R_1(x_1,x_2) > 1$, and some extra precautions would be necessary.)  
The correction factor $R_2$ for the other shower history looks the same. 
Put another way, the relative probability for a gluon to be emitted by 
parton 1 or by parton 2 is unchanged by the matrix element correction 
procedure. 

The above procedure works not only for quarks, but also e.g. for squarks.
The numerator of the splitting kernel would have been $2z$ rather than
$1+ z^2$ \cite{stoprad}, but both are equally well approximated from above 
by 2. For a gluino, the colour charge is $N_C$ rather than $C_F$, so the 
assumed shower emission rate has to be scaled up by a factor 
$N_C/C_F = 9/4$ (also for the recoiling colour triplet parton, since the 
separation of radiation is not perfect), with no other change required. 

We will also encounter processes where the decaying particle carries colour,
like in $\t \to \b \W^+$. Then gluon emission off this particle has to 
be considered, i.e. graphs like
$\t(0) \to \t(i) \, \g(3) \to \b(1) \, \W^+(2) \, \g(3)$,
proceeding via an intermediate off-shell top. Neglecting the top width, this 
introduces a new kind of inverse propagators
\begin{equation}
|Q_0^2| = m_0^2 - m_i^2 = m_{123}^2 - m_{12}^2 = 2 (p_1 + p_2) p_3 
= x_3 \, \ECM^2 ~.    
\end{equation}
The leading term of the matrix element, proportional to $1/(Q_0^2Q_1^2)$, will
therefore have the same singularity structure as the shower rate in
eq.~(\ref{PSfirsthist}). In a shower description, where only the $\b$ is allowed
to radiate, there is thus no problem in principle of letting that radiation
account for the full emission pattern of the matrix elements, i.e.
\begin{eqnarray}
R_1(x_1,x_2) & = &
\frac{W_{\mathrm{ME}}(x_1,x_2)}{W_{\mathrm{PS,1}}(x_1,x_2)} 
\nonumber \\
& = & \frac{x_3(1 + r_2^2 - r_1^2 - x_2)}{2} \,
\left( \frac{\alphas}{2\pi} \, C_F \right)^{-1} 
\frac{1}{\sigma_0} \, \frac{\d \sigma_{\mathrm{ME}}}{\d x_1 \, \d x_2}~.
\label{massivecorr2}
\end{eqnarray}
Since also here the shower rate turns out to exceed the matrix element one,
a simple rejection approach should work well.

\subsection{Subsequent branchings}

So far, we have considered matrix elements as providing the probability of
exclusive three-jet events. An alternative interpretation, however, would be
in terms of an inclusive density of gluon emissions, with the possibility
of several such per event. This interpretation works well in the soft-gluon 
limit, while the emission of a hard gluon reduces the phase space for
subsequent emissions and thus ruins the picture of independent emissions.
Furthermore, the possibility for gluons to branch in their turn leads to the
need to include coherence effects \cite{coherence} that constrain allowed
emissions. Nevertheless, the matrix elements can be used to extract
important information, not only on the emission rate of hard gluons, but 
also on that of soft and collinear ones. 

Historically, different approaches have therefore been taken.
In the \textsc{Jetset/Pythia} procedure used until now, only the first 
branching is corrected by matrix-element information. Subsequent emissions 
involve no corrections, but only the process-independent splitting
kernels, and therefore give the wrong emission rate off
heavy quarks that has already been noted. In the \textsc{Herwig} routine,
a correction is performed for every emission that is the hardest so far 
\cite{Mikecorr} --- whereas \textsc{Jetset} emissions tend to be ordered 
in hardness, this is less likely in \textsc{Herwig}, so the distinction is 
then relevant. \textsc{Ariadne}, finally, imposes a correction at all steps 
of the cascade.

Since our new algorithm does not have the correct dead-cone behaviour 
built in from the onset, it is clear that some correction procedure 
will be required. Rather than imposing the process-independent collinear 
behaviour, we have chosen to base ourselves on the matrix elements also 
in this region. This offers a smooth interpolation between the hard 
process-specific and the soft collinear universal behaviours, and sidesteps 
the issue of which emissions are to be considered the hardest so far
\cite{Mikecorr}. Thus the shower rate will be corrected to the matrix element one at 
every step of the shower off the original partons. The gluon cascading 
$\g \to \g\g$ is unaffected, since there are no gluon-mass effects or
matrix elements to be considered here. Light quarks (in 
$\gamma^*/\Z^0$ decay) are also essentially unaffected, since there the
shower matches so well with the matrix elements anyway.

The kinematics of the cascade changes in each emission, as the energy of the 
radiating parton is reduced by the previous emissions. The mapping of an
emission on to the matrix element variables is thereby not unique. However,
from the dead cone formula, eq.~(\ref{deadconesuppr}), we see that the 
emission angle and the mass-to-energy ratio of the emitting quark should
be represented faithfully in the choice of matching matrix element variables.
For the branchings subsequent to the first one, say $3 \to 7 + 8$ in 
Fig.~\ref{fig:Zshower}a, the mother energy $E_3$ is fixed. The choice of a 
$Q^2 = m_3^2 - m_7^2$ and a $z$ of the branching maps onto four-momenta $p_7$ 
and $p_8$ as described in eqs.~(\ref{shuffle})--(\ref{pTbranching}), with  
$p_7^2 = m_{\q}^2 > 0$ and $p_8^2 = 0$. In order to make contact with the 
matrix element variables, now construct a hypothetical recoiling parton $2'$,
such that $\mathbf{p}'_2 = - \mathbf{p}_3$ and ${p'_2}^2 = m_{\qbar}^2$.
The CM energy of this reduced system is then given by
\begin{equation}
\EPCM = E_3 + E'_2 = E_3 + \sqrt{\mathbf{p}_3^2 + m_{\qbar}^2} =
E_3 + \sqrt{E_3^2 - Q^2 - m_{\q}^2 + m_{\qbar}^2}  ~.
\end{equation}
Now matrix element and parton shower weights can be evaluated for
$x'_1 = 2 E_7/\EPCM$, $x'_2 = 2 E'_2/\EPCM$, $r'_1 = m_{\q}/\EPCM$ and
$r'_2 = m_{\qbar}/\EPCM$. Note that, in the limit $E_4 \to 0$, also
$\EPCM \to \ECM$, and the ordinary matrix elements are recovered.

The above procedure is not unique. As an alternative, one could retain
the original $\ECM$, but construct an off-shell $p'_2$ to carry all the 
energy and momentum of the system except for $p_3 = p_7 + p_8$, 
i.e. $p'_2 = p_2 + p_4$ in the case of Fig.~\ref{fig:Zshower}a. This gives
equivalent results so long as the gluon energy already emitted is not too 
large, and else gives a somewhat lower rate of wide-angle emission, 
reflecting that the parton $2'$ then is assigned such a large mass that it
radiates less. The overall picture therefore is less appealing, even if
results in practical applications are almost equivalent. 

In this article, emphasis is put on the gluon emission off the
primary quarks (or other primary particles). The subsequent branchings
$\g\to\g\g$ have not been affected by the mass considerations, so
are not discussed here. The rate of $\g \to \q\qbar$ branchings is 
another topic of some interest, given that LEP results do not quite
agree with predictions \cite{QCDWG}. This issue is further discussed 
in Sect.~6. 

However, given that such a secondary $\c\cbar$ or $\b\bbar$ 
pair has been produced, the possibility of further gluon emission off 
this pair should be considered, even if most pairs are produced at
such a low mass that the phase space left for further
radiation is limited. In order to provide a sensible 
behaviour in the collinear region, again matrix-element input is
applied, calculated for the decay of a colour octet source. 
(Process 66 of Table~\ref{tab:processes}, so with the wrong spin of 
the source, but correct for the radiating parton, which is the main 
point.) To first approximation, this means that radiation occurs 
independently off the $\q$ and $\qbar$; see the discussion on
radiation patterns below. The kinematics for the matrix element
corrections is set up about as described above, i.e. by mapping
onto a reduced system at rest with preserved energy for the radiating 
parton. A simplification is that here the recoiling parton always
has the same mass as the radiating one. 

\subsection{Additional issues}
\label{sec:additional}

In the older shower algorithm, the decay products of a branching 
$a\to bc$ were assumed massless until assigned a mass by some 
subsequent step. This meant that the $\pT^2$ of a branching, 
eq.~(\ref{pTbranching}) with $\beta_a =1$ as a first simplification, 
was simplified further to $\pT^2 \approx z (1-z) m_a^2$ when used 
as argument in $\alphas(\pT^2)$. In a branching $\q\to\q\g$, where 
the $\q$ is assumed to have a non-vanishing rest mass, one would
instead obtain $\pT^2 \approx z (1-z) m_a^2 (1-m_b^2/m_a^2)^2$.
(This follows from eq.~(\ref{pTbranching}) or alternatively from
the rescaling of $x_3$ in eq.~(\ref{shuffleqqg}).) A smaller 
$\pT^2$ estimate implies a larger $\alphas$ and emission rate  
for a given kinematical configuration, but also that more phase
space is cut out by the requirement $\pT^2 > Q_0^2/4$, where $Q_0$
is the soft cut-off scale of the cascade. The net result of such a
change is therefore not obvious, and we will study it later on.

Also the calculation of the approximate opening angle of a branching
$a\to bc$, eq.~(\ref{angle}), would be affected by the same 
considerations. The massive parton energy $E_b$ is increased at
the expense of $E_c$, as given by eq.~(\ref{shuffleqqg}). Since the 
common $\pT$ is also decreased, the ratio $\pT/E_c$ is preserved, 
while $\pT/E_b$ is decreased. The net result is a decrease of the
opening angle by a factor $(1 + (m_b^2/m_a^2)(1-z)/z)^{-1}$.
As another option, we will then consider the consequences of such a
decrease in the decay opening angle, without any change of the 
production angle, as a way of minimally relaxing the angular ordering 
condition off massive quarks.

While our choice of $Q^2 \approx m^2$ variable has significant
advantages for the matching to matrix-element expressions,
it does not offer as neat an implementation of coherence effects
as the angular variable of {\sc Herwig} or the transverse momentum
one of {\sc Ariadne}. Without any further constraints, the amount
of radiation is overestimated. Therefore, by default, angular
ordering of emissions is imposed as a further constraint. This,
on the other hand, tends to restrict emissions somewhat too much.
The bulk of these ambiguities affect rather soft gluons, which do
not give rise to separate jets. For the precision studies in
Section~\ref{sec:bottom}, however, also small effects could be
of interest. We have therefore introduced a new ``intermediate''
coherence option, as compared to the minimal modification above,
wherein no angular constraint is imposed on emissions
off the primary $\q\qbar$ pair. These emissions thus are ordered
only in mass. Angular ordering is still imposed in the cascades initiated
by the gluons emitted off the primary quarks.
In particular each such cascade is restricted
to a cone given by the emission angle of the initiating gluon.
In this option, generic event properties are only slightly changed
compared with the default procedure.

So far, we have mainly considered configurations where a colour
singlet decays to stable particles, e.g. $\gamma^*/\Z^0 \to \b\bbar$.
Another class of events involve sequential decays of coloured objects.
The obvious example would be 
$\gamma^*/\Z^* \to \t\tbar \to \b\W^+\bbar\W^-$. Even leaving aside 
the continued fate of the $\W$'s, e.g. assuming they decay leptonically,
the event now contains four colour charges that may radiate. In the
limit $\Gamma_{\t} \to 0$, the radiation in the top production stage
$\gamma^*/\Z^* \to \t\tbar$ decouples completely from that in the top
decays. For a finite $\Gamma_{\t}$, gluons with energies below or around 
this scale can receive contributions from many colour sources, however,
leading to complex radiation patterns \cite{widtheffect}. Close to
the $\t\tbar$ threshold, the main sources are the respective top decays,
described by dipoles $\widehat{\t\b}$ and $\widehat{\tbar\bbar}$,
and radiation off the $\widehat{\b\bbar}$ dipole created after the $\t$
and $\tbar$ have decayed. Gluons with energies above (below) 
$\Gamma_{\t}$ predominantly feel the former (latter) dipoles. The 
radiation pattern can be written as
\begin{equation}
\frac{E_{\g}^2}{\Gamma_{\t}^2 + E_{\g}^2} 
\left( \widehat{\t\b} + \widehat{\tbar\bbar} \right) +
\frac{\Gamma_{\t}^2}{\Gamma_{\t}^2 + E_{\g}^2} 
\widehat{\b\bbar} ~.
\label{Gammadampen}
\end{equation}
Thus the reduced radiation induced by the top finite lifetime
is compensated by the radiation from new sources that would not 
have been present for a long-lived top. The $\widehat{\b\bbar}$ dipole
introduces a dependence on the opening angle between the $\b$ and
the $\bbar$ that is not there in the separate top decays,
so the compensation is not complete. On the perturbative level,
this gives a dipole effect \cite{Leningraddipole} that could be 
observed at low momenta. However, even in the approximation of allowing
radiation only within each top quark decay separately, the string 
fragmentation picture \cite{Lundstring} would imply that a 
nonperturbative colour string should be stretched between the 
$\b$ and $\bbar$ (or their respective cascades), and this would
introduce a string effect \cite{stringeff} of almost equal character 
and magnitude \cite{topstudy}. Thus only very careful studies, e.g. 
for high-precision measurements of the top mass, would be sensitive
to the detailed nature of the soft-gluon emission source.
The critical transition is the one between a top long-lived enough 
to produce top hadrons and one decaying too rapidly for that, where
the hadronic final state does change character. 

The shower algorithm contains options that allows it to be run,
either with soft or with hard emission dampened according to the
respective factors in eq.~(\ref{Gammadampen}). However, currently the
\textsc{Pythia} program contains no machinery to detect when
a description of this kind is required, nor a prescription how to 
combine all possible sources of radiation. This is obviously an
interesting task for the future, but one that will be required 
primarily for particles with a width significantly larger than that 
of the top. 

The normal generation sequence therefore contains a set of 
separated showers. For instance, in the top example above, the
$\gamma^*/\Z^* \to \t\tbar$ induces a first cascade, whereby 
the $\t$ and $\tbar$ shower down to the mass shell. Thereafter
follows the respective top decay, $\t \to \b \W^+$ and
$\tbar \to \bbar\W^-$, and the separate radiation in those decays.
At some yet later stage, the two $\W$'s may decay to quarks that
again radiate. Anytime a colour singlet is exchanged, like the 
$\W$'s above, there is a clean separation into disjoint QCD 
subsystems, while the exchange of a coloured state like the top
will hook up separately showering systems to the strings that
later will produce the observable hadrons. Interconnection effects
\cite{interconnection} could complicate this picture, but so far
the evidence is that any such effects would be small. 

In hadronic collisions, more complex processes would occur,
and also initial-state QCD radiation has to be considered as 
a potential source of further interference effects. 
In this article we will not address these additional complications,
but defer that for some future study. Currently such 
interference is almost completely neglected, except for some 
angular restrictions \cite{CDF_ISRinterference}.

Even if $\Gamma_{\t}$ is neglected in the showering, for the sake of 
providing a unique separation of radiation before and after the top 
propagator, this does not mean that all top quarks have to have the 
same mass. Instead, resonance masses are chosen according to the 
relevant Breit-Wigners, convoluted with the respective cross section 
formulae. In particular, this means that the $\t$ and $\tbar$ masses 
of an event would be unequal. The shower algorithm described above
then operates on these event-by-event masses, without any reference
to their nominal on-shell equivalents.

The hadronization of a partonic configuration, obtained by the chain
of decays and showers outlined above, is described by the Lund string 
model \cite{Lundstring}. All coloured partons belong to strings,
stretched from a quark endpoint via a number of intermediate gluons
to an antiquark one. (Also diquark endpoints and closed gluon loops 
can be considered, but are of no relevance in this article.) Normally
each such string would have a reasonably large invariant mass, enough
to produce several hadrons. However, occasionally a string could come 
to have a small mass, e.g. by the splitting of a string in 
two by shower branchings $\g\to\q\qbar$. Then a special treatment may
be required for one- or two-hadron decays. While the normal string
algorithm has remained essentially unchanged over a number of years,
this low-mass `cluster' treatment has recently been improved 
\cite{clusterstring}. The direct consequences for the topics studied
here are minimal, except that quark mass values have now been 
optimized, especially to describe charm asymmetries in fixed-target
experiments. The current default values thus are 
$m_{\u} = m_{\d} = 0.33$~GeV, $m_{\s} = 0.5$~GeV,
$m_{\c} = 1.5$~GeV and $m_{\b} = 4.8$~GeV.

\section{Matrix elements}

As input and starting point for our shower studies, we will need the matrix 
elements for the processes of interest. In this article this essentially 
means the two-body decay of a particle, with associated gluon radiation.
Some of the formulae are available in the literature, but most are not, 
or at least not easily found. We have therefore calculated a number of 
processes. This also gives us a chance to test the degree of universality
of the radiation patterns in channels of different colour and spin structure,
but with the same masses. These results are interesting in their own right.

\begin{table}[t]
\begin{center}
\begin{tabular}{|c|c|c|c|c|@{\protect\rule[-2mm]{0mm}{7mm}}}
\hline
colour & spin & $\gamma_5$ & example & codes \\
\hline
$1 \to 3 + \overline{3}$ & --- & --- & (eikonal) & 6 -- 9 \\
$1 \to 3 + \overline{3}$ & $1 \to \frac{1}{2} + \frac{1}{2}$ & 
$1,\gamma_5,1\pm\gamma_5$ & $\Z^0 \to \q\qbar$ & 11 -- 14 \\
$3 \to 3 + 1$ & $\frac{1}{2} \to \frac{1}{2} + 1$ & 
$1,\gamma_5,1\pm\gamma_5$ & $\t \to \b\W^+$ & 16 -- 19 \\
$1 \to 3 + \overline{3}$ & $0 \to \frac{1}{2} + \frac{1}{2}$ & 
$1,\gamma_5,1\pm\gamma_5$ & $\H^0 \to \q\qbar$ & 21 -- 24 \\
$3 \to 3 + 1$ & $\frac{1}{2} \to \frac{1}{2} + 0$ & 
$1,\gamma_5,1\pm\gamma_5$ & $\t \to \b\H^+$ & 26 -- 29 \\
$1 \to 3 + \overline{3}$ & $1 \to 0 + 0$ & 
$1$ & $\Z^0 \to \sq\sqbar$ & 31 -- 34 \\
$3 \to 3 + 1$ & $0 \to 0 + 1$ & 
$1$ & $\sq \to \sq'\W^+$ & 36 -- 39 \\
$1 \to 3 + \overline{3}$ & $0 \to 0 + 0$ & 
$1$ & $\H^0 \to \sq\sqbar$ & 41 -- 44 \\
$3 \to 3 + 1$ & $0 \to 0 + 0$ & 
$1$ & $\sq \to \sq'\H^+$ & 46 -- 49 \\
$1 \to 3 + \overline{3}$ & $\frac{1}{2} \to \frac{1}{2} + 0$ & 
$1,\gamma_5,1\pm\gamma_5$ & $\tilde{\chi} \to \q\sqbar$ & 51 -- 54 \\
$3 \to 3 + 1$ & $0 \to \frac{1}{2} + \frac{1}{2}$ & 
$1,\gamma_5,1\pm\gamma_5$ & $\sq \to \q\tilde{\chi}$ & 56 -- 59 \\
$3 \to 3 + 1$ & $\frac{1}{2} \to 0 + \frac{1}{2}$ & 
$1,\gamma_5,1\pm\gamma_5$ & $\t \to \st\tilde{\chi}$ & 61 -- 64 \\
$8 \to 3 + \overline{3}$ & $\frac{1}{2} \to \frac{1}{2} + 0$ & 
$1,\gamma_5,1\pm\gamma_5$ & $\sg \to \q\sqbar$ & 66 -- 69 \\
$3 \to 3 + 8$ & $0 \to \frac{1}{2} + \frac{1}{2}$ & 
$1,\gamma_5,1\pm\gamma_5$ & $\sq \to \q\sg$ & 71 -- 74 \\
$3 \to 3 + 8$ & $\frac{1}{2} \to 0 + \frac{1}{2}$ & 
$1,\gamma_5,1\pm\gamma_5$ & $\t \to \st\sg$ & 76 -- 79 \\
$1 \to 8 + 8$ & --- & --- & (eikonal) & 81 -- 84 \\
\hline
\end{tabular}
\caption{The processes that have been calculated, also with one
extra gluon in the final state. Colour is given with 1 for singlet, 
3 for triplet and 8 for octet. See the text for an explanation of 
the $\gamma_5$ column and further comments.}
\label{tab:processes}
\end{center}
\end{table}

\subsection{Calculations}
\label{sec:calc}

A number of matrix elements have been calculated, using \textsc{Comphep}
\cite{Comphep} for the actual calculation, including an extension
package for Supersymmetric processes \cite{Comphepsusy}, and 
\textsc{Mathematica} \cite{Mathematica} for subsequent simplification 
of the expressions. The list of lowest-order (LO) processes is given in 
Table~\ref{tab:processes}. The LO expression gives the two-body decay rate of
a particle, i.e. $a \to b c$, and the matching first-order (FO) one the same 
decay with an additional gluon in the final state, $a \to b c \g$.

While the matrix element calculations in this section have been 
performed from scratch, some checks are based on results in the 
literature. These include: 
$V \to \q\qbar$ for $m_{\q} = m_{\qbar}$ \cite{MEqqgmassive},
$\H^0 \to \q\qbar$ for $m_{\q} = m_{\qbar}$ \cite{Janot},
$V \to \sq\sqbar$ for $m_{\sq} = m_{\sqbar}$ \cite{stoprad},
and $\t \to \b\W^+$ for $m_{\b} = 0$ \cite{Miketdec}. 
No doubt, many more are available, without our knowledge.

The process selection is based on the particle content of the
Minimal Supersymmetric Standard Model (MSSM), i.e. includes squarks
$\sq$, gluinos $\sg$, neutralinos and charginos $\tilde{\chi}$,
and Higgs states
$\hrm^0$, $\H^0$, $\A^0$ and $\H^{\pm}$. The idea, however, is that
these situations could represent also a number of other non-standard
particles. For instance, the decay of a spin 0 leptoquark 
$\mathrm{L}_{\mathrm{Q}} \to \q\ell$ is closely similar to
$\sq \to \q\tilde{\chi}$.

All calculations have been performed in the zero-width limit of the 
decaying state and the decay products, in order to allow a gauge 
invariant separation of radiation in the production and decay stages.
(As explained above, the Monte Carlo simulation of processes does 
include mass selection according to the appropriate Breit-Wigners, so 
we here only comment on the width dependence of the additional gluon 
radiation, i.e. in the ratio of first to leading order cross sections.)   
Such a separation occurs naturally for exchanged colourless particles,
so is then no problem. For a coloured particle of width $\Gamma$, this 
is a poor approximation in the region of gluon energies (in the particle 
rest frame) below $\Gamma$. The $\t$ is still sufficiently narrow that 
the production--decay interference is not a major problem, see above, 
but for heavier squarks and gluinos a more complex description may be 
required. The lowest coloured SUSY states, e.g. stop, tend to have small 
widths, however, and it is on such states we will concentrate our 
studies. The additional complications for very wide particles will be
deferred to some future study. 

The classification by colour and spin is fairly obvious, but that does
not completely specify the structure of the process. Consider e.g.
$\e^+\e^- \to \gamma^*/\Z^0 \to \q \qbar$. The cross section 
(neglecting mass effects) is then
\begin{equation}
\sigma_0 \propto e_{\e}^2 \, e_{\q}^2 + 
2 \, e_{\e}v_{\e} \, {\Re}\mathrm{e} \Xi \, e_{\q}v_{\q} +
(v_{\e}^2 + a_{\e}^2) \, |\Xi|^2 \, (v_{\q}^2 + a_{\q}^2) ~,
\label{gamZmix}
\end{equation} 
where 
\begin{equation}
\Xi = \frac{1}{16 \sin^2\theta_W \cos^2\theta_W} \,
\frac{s}{s - m_{\Z}^2 + i m_{\Z} \Gamma_{\Z}}
\end{equation} 
represents the ratio between the $\Z^0$ and the $\gamma$ propagators and
couplings. The term proportional to $a_{\q}^2$ corresponds to the 
$\q\qbar$ pair coming from an axial vector source, the rest to it
coming from a vector source. Since the QCD radiation from these two is
somewhat different, we need to include the proper mixture, which depends
on the CM energy, $s = \ECM^2$. It has then been simplest to perform
the calculations for a pure vector source and a pure axial vector source, 
represented by 1 and $\gamma_5$ in Table~\ref{tab:processes}, and mix in 
the proportions required. We note that this mixing strategy is possible 
since the total cross section does not contain any interference
terms of the character $v_{\q} a_{\q}$. (Such terms do arise when the 
forward--backward charge asymmetry is considered.) The pure left-handed
mixture $V - A = 1 - \gamma_5$ is of special interest, since it represents
the $\W^{\pm}$ bosons. Therefore it has been calculated separately,
and is denoted by $1\pm\gamma_5$ in Table~\ref{tab:processes}; as we 
already noted the sign (of the interference term in the squared matrix 
element) is irrelevant for the QCD emission aspects. In total, four 
alternatives are therefore open in our implementation: to have a pure 
vector source, a pure axial vector, an arbitrary mixture 
$\alpha V + (1 - \alpha) A$ and the special equal mixture. In this 
order, that gives the four codes 11 -- 14 in Table~\ref{tab:processes}, 
according to the numbering scheme used in the \textsc{Pythia} function
\texttt{PYMAEL} introduced in version 6.153.
(This routine takes as input the process code, $x_1$, $x_2$,
$r_1$, $r_2$ and $\alpha$, and returns the ratio 
$(1/\sigma_0) \, \d \sigma / \d x_1 \, \d x_2$, omitting a factor of
$(\alphas/2\pi) C_F$.) 

Correspondingly, also most other processes can come either with or without
$\gamma_5$ factors in the amplitude, or in arbitrary mixtures thereof. 
In the Higgs sector, normally the $\hrm^0$ and $\H^0$ are scalar, the 
$\A^0$ pseudoscalar and the $\H^{\pm}$ a parameter-dependent mixture of 
1 and $\gamma_5$. If the coupling structure is
generalized, also the neutral Higgses could be mixtures, however. 
The $\sq_L$ and $\sq_R$ squark partners of the left- and right-handed quarks 
come with wave function factors $1-\gamma_5$ and $1+\gamma_5$, 
respectively. The squark mass eigenstates will be mixtures of these, 
with significant mixing expected 
especially in the third generation. With two squarks in a process,
$\gamma_5^2 = 1$ ensures that the matrix element still can be written 
as a sum $\alpha \, 1 + (1 -\alpha) \gamma_5$. Again, therefore, we have 
chosen to perform most of the calculations with and without a $\gamma_5$ 
factor and then leave open to have the mixing depend on the current parameter 
choice. As a further simplification, the $1\pm\gamma_5$ mixture is used 
whenever the correct choice is not known, since this mixture represents 
an average behaviour. In some instances, further restrictions exist,
e.g. a pseudoscalar cannot decay to two scalars, at least among the MSSM
processes at our disposal.

The process at the top of Table~\ref{tab:processes} is the 
spin-independent eikonal answer of eq.~(\ref{eikonalexpr}), extended 
from the soft region where it is intended to be valid: 
\begin{eqnarray}
\frac{1}{\sigma_0} \, \frac{\d \sigma}{\d x_1 \, \d x_2}
& = & \frac{\alphas}{2\pi} \, C_F  \left\{ 
\frac{2 ( x_1 + x_2 -1 -r_1^2 -r_2^2)}{(1 + r_1^2 - r_2^2 - x_1)%
(1 + r_2^2 - r_1^2 - x_2)} \right. \nonumber \\
& & \left. - \frac{2 r_1^2}{(1 + r_2^2 - r_1^2 - x_2)^2} -
\frac{2 r_2^2}{(1 + r_1^2 - r_2^2 - x_1)^2} \right\} ~.
\label{eikonalxx}
\end{eqnarray}
The first numerator, $2 ( x_1 + x_2 -1 -r_1^2 -r_2^2)$, here is based on
an evaluation of the $2p_1p_2$ numerator of eq.~(\ref{eikonalexpr})
with $p_1$ and $p_2$ given by their values after the emission of $p_3$.
Away from the soft-gluon limit, however, there is some leeway in this 
assumption. If instead the $p_1$ and $p_2$ values before the emission 
of $p_3$ had been used, one would have obtained 
$x_1 + x_2 -1 = 1 - x_3 \to 1$. Such a seemingly minor substitution has
quite dramatic effects for collinear emission even at rather small
$x_3$, however, and is not really an option. In order to have a not too 
unrealistic alternative to compare with, we therefore only allow 
deviations proportional to $x_3^2$. (This is also what comes out of 
the $x_1^2 + x_2^2$ numerator of the massless process 
$V \to \q\qbar\g$.)  The extreme in this direction would 
be $x_1 + x_2 -1 \to 1 - x_3 + x_3^2$, but we will also allow
arbitrary admixtures $x_1 + x_2 -1 \to 1 - x_3 + \alpha x_3^2$,
with $\alpha$ as a free parameter, obviously in no physics relation 
to the $\alpha$ introduced above.

In all the formulae, the decay product mass ratios are kept as free 
parameters, $r_1 = m_b/m_a$ and $r_2 = m_c/m_a$, while the gluon is 
massless. Since we are not interested in cross sections per se, but 
in the probability for gluon emission, in the LO cross sections only 
the mass dependence is retained, normalized to unity for $r_1 = r_2 = 0$,
e.g.
\begin{equation}
\sigma_0 (V,A \to \q_1\qbar_2) =
\frac{1}{2} \left\{ 2 - (r_1^2 - r_2^2)^2 - r_1^2 - r_2^2 \pm 6 r_1 r_2
\right\} \sqrt{(1 -r_1^2 -r_2^2)^2 - 4 r_1^2 r_2^2} ~.
\label{VAmass}
\end{equation}
What is omitted is then some set of couplings and propagators.
Exactly the same set is also omitted from the first-order 
$\d \sigma / \d x_1 \, \d x_2$, thereby leaving the ratio 
unchanged. Additionally the common factor $(\alphas/2\pi) C_F$, is omitted, 
to leave the choice of $\alphas(\pT^2)$ free to be made elsewhere. 
The ratio of the first to leading order cross sections then gives the
assumed differential gluon-emission rate. In the case of a mixture without
and with $\gamma_5$ factors, the sum has to be taken for numerator and
denominator separately, e.g.
\begin{equation}
\frac{1}{\sigma_0} \, \frac{\d \sigma}{\d x_1 \, \d x_2} 
(\e^+e^- \to \gamma^*/\Z^0 \to \q \qbar) =
\frac{\alpha \, \d \sigma^V/\d x_1 \, \d x_2 +
(1-\alpha) \, \d \sigma^A/\d x_1 \, \d x_2}%
{\alpha \, \sigma_0^V + (1-\alpha) \, \sigma_0^A} ~.
\label{sigratgamZ}
\end{equation}
The vector fraction $\alpha$ can here be read off from the lowest-order
expression in eq.~(\ref{gamZmix}), where the mass factors of  
eq.~(\ref{VAmass}) were omitted. The mass effects are instead 
included in the individual $\d \sigma/\d x_1 \, \d x_2$ and $\sigma_0$
terms of eq.~(\ref{sigratgamZ}). This standard e.g. means that 
$V - A$ corresponds to $\alpha = 1/2$, which then because of mass 
effects gives a somewhat larger vector than axial fraction.

Note that first-order corrections to the total cross section are not included in the 
$\sigma_0$ denominator. This is not a unique choice, but a 
rather natural one: if we consider the gluon emission rate as the ratio 
of two cross sections, then including $\mathcal{O}(\alphas)$ corrections
to one but not the other is not likely to improve the overall accuracy of 
the calculation. In the soft-gluon limit, it would even break the
spin independence of the radiation pattern (cf. the next subsection),
i.e. give the wrong physics. And since a complete one-loop calculation 
of both quantities is well beyond the scope of this article, we remain 
with lowest non-trivial order for both quantities. Furthermore, if the 
total cross section is written in the form 
$\sigma_{\mathrm{tot}} = \sigma_0 (1 + a \alphas/\pi)$,
then decay rates calculated so far tend to give $a$ values of order
unity \cite{Janot,ascorrtoLO}, i.e. small effects. An exception would 
be Coulomb corrections in the threshold region, but there the phase 
space for real gluon emission is vanishingly small anyway, so of no 
physical interest. By contrast, one-loop corrections to three-jet rates 
tend to be larger \cite{ERT}, although most of that is absorbed by the 
parton-shower choice of a smaller kinematics-dependent scale like 
$\pT^2$ \cite{ERT, Kunszt}.

\subsection{Radiation patterns}
\label{sec:radpat}

\begin{figure}[tp]
\begin{center}
\epsfig{file=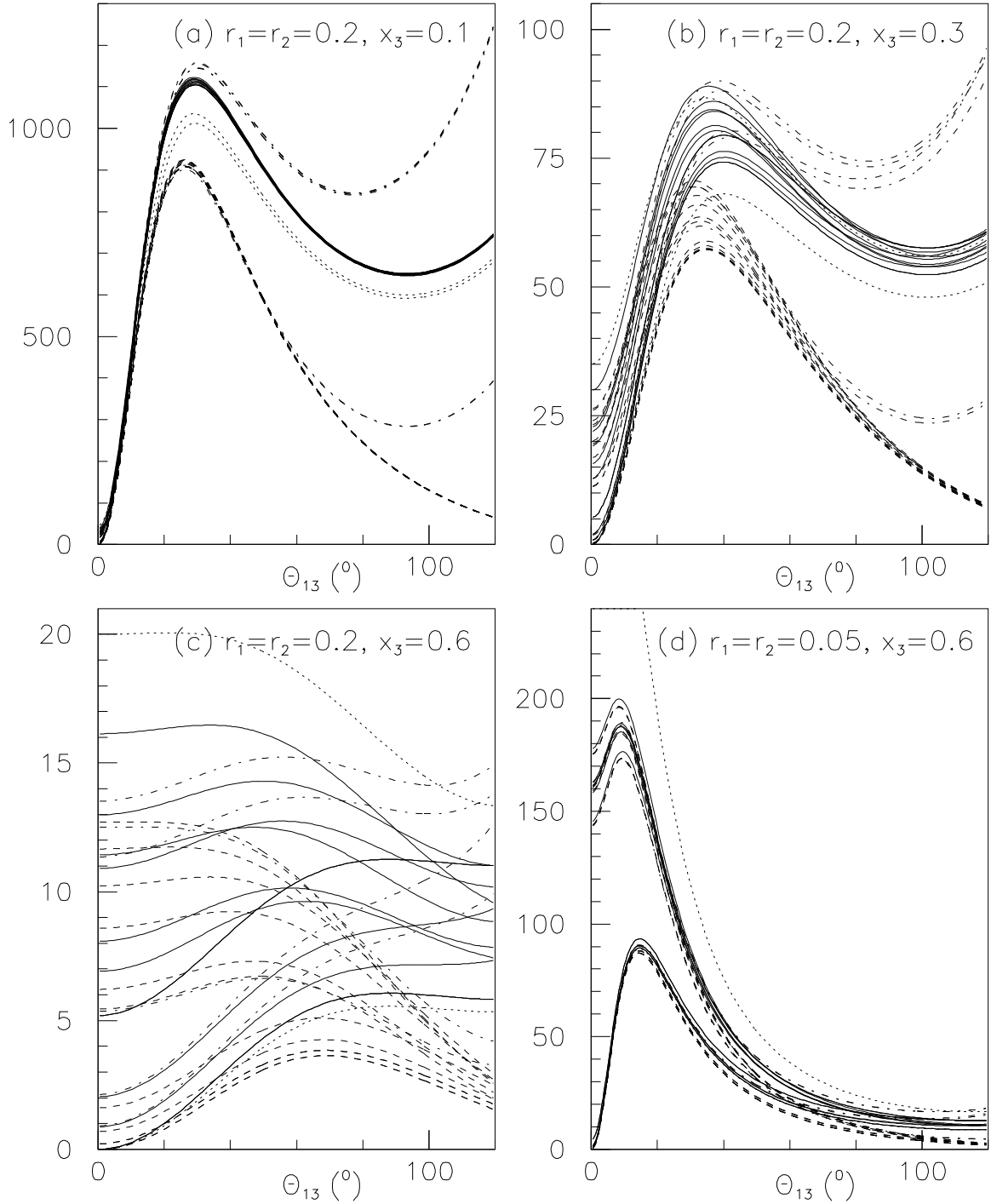}
\end{center}
\vspace{-2\baselineskip}
\caption{The gluon emission rate as a function of the 
emission angle $\theta_{13}$. Specifically, the vertical axis
gives $(1/\sigma_0) \d \sigma / \d x_1 \, \d x_2$ with a normalization factor
$(\alphas/2\pi)C_F$ removed. This three-jet phase 
space density differs from $\d \sigma / \d \theta_{13} \, \d x_3$
by a simple Jacobian. A variety of different processes are 
compared: decay of a colour singlet to a triplet plus an 
antitriplet full curves, ditto in the eikonal approximation 
dotted, decay of a triplet to a triplet plus singlet dashed, 
and gluino processes dash-dotted. The four frames differ in the 
scaled masses $r_i = m_i/\ECM$ of the two decay products, and in 
the gluon energy fraction $x_3$. Further explanations are given 
in the text.}
\label{fig:compproc}
\end{figure}

Before applying the matrix elements to specific physics situations, 
it is interesting to compare them between each other under
similar conditions. This will provide an understanding of the amount 
of spin and colour dependence in the matrix elements, and thus the
extent to which a process-dependent Monte Carlo implementation
can be expected to provide an improvement over a process-blind
one, e.g. based on some simple dead cone formula.  

Fig.~\ref{fig:compproc} is intended to provide a first qualitative
glimpse of differences. The amount of detail may be bewildering, but 
at this point the idea is to bring up differences and similarities
in a broad sense, without concentrating on each process specifically.

First consider the full curves in Figs.~\ref{fig:compproc}a--c,
which are all for the colour structure $1 \to 3 + \overline{3}$
and the same kinematics, and only differ by the spin pattern:
$1 \to 1/2 + 1/2$, $0 \to 1/2 + 1/2$, $1 \to 0 + 0$, $0 \to 0 + 0$,
and $1/2 \to 1/2 + 0$, in relevant cases with and without a $\gamma_5$
(but no intermediate mixtures such as $1\pm\gamma_5$).
In the first frame, where the gluon is soft, these curves almost
completely overlap. At smaller $x_3$ the agreement becomes even
better, and one can then truly speak of a universal soft-gluon 
emission pattern, with a characteristic dead cone of opening 
angle $\theta \approx 2r = 0.4 = 23^{\circ}$ in this example. By contrast, when 
$x_3$ is increased, the curves tend to disagree more and more.
At $x_3 = 0.3$, the dead cone is still visible, although it is
starting to fill in, but at  $x_3 = 0.6$ it is completely gone.
Note that the vertical scale changes significantly between the 
three frames; in absolute numbers the differences between the 
curves is about a factor two larger at $x_3=0.1$ than at 0.6, 
but in relative terms the differences are negligible at small $x_3$. 

This should bring home the message that the dead cone concept
can only be used at small gluon energies, and is irrelevant
for energetic gluons. It is only the lowest of the full 
curves, the completely spinless process $0\to 0 + 0$, that does preserve
the exact dead cone concept, i.e. has a cross section that always
vanishes in the collinear limit (in the rest frame of the decaying
particle). 

The lower dotted curve is the eikonal expression of 
eq.~(\ref{eikonalxx}), which is constructed to have an exact dead 
cone. It does agree fairly well with the spinless process, but not 
with anything else. The upper dotted curve is the modified eikonal,
with an $x_3^2$ term added. It tends to overshoot all processes at
small angles, while some intermediate mixture, $\alpha \approx 0.5$,
would do a sensible job for many processes in the small-angle region.
For medium small gluon energies and large angles, both eikonal forms 
undershoot, however. In general, one may conclude that the eikonal 
formula is not particularly useful for practical considerations, and 
that the process-specific matrix elements need to be used.

Of course, the detailed pattern depends on the masses. In 
Fig.~\ref{fig:compproc}d these are lower than in the first three, and 
representative for $\Z^0 \to \b\bbar$. Then the dead cone concept works  
up to somewhat larger gluon energies, and a trace of it is still left
at $x_3 = 0.6$. It is again preserved exactly for $0\to 0 + 0$ and
the eikonal, and also approximately for $1 \to 0 + 0$.  

\begin{table}[t]
\begin{center}
\begin{tabular}{|c|c|c|c|c|c|@{\protect\rule[-2mm]{0mm}{7mm}}}
\hline
colour & spin & $\gamma_5$ & $\int x_3$ & 
$\int (1-x_1)(1-x_2)$ & $\int \theta(y_{\mathrm{D}}-0.1)$ \\
\hline
$1 \to 3 + \overline{3}$ & $1 \to \frac{1}{2} + \frac{1}{2}$ & 1 &  
1.000 & 1.000 & 1.000 \\
 & & $\gamma_5$ & 1.056 & 1.112 & 1.133 \\
 & $0 \to \frac{1}{2} + \frac{1}{2}$ & 1 & 1.134 & 1.293 & 1.376 \\
 & & $\gamma_5$ & 1.093 & 1.207 & 1.271 \\
 & $1 \to 0 + 0$ & 1 & 1.073 & 1.205 & 1.310\\
 & $0 \to 0 + 0$ & 1 & 0.875 & 0.758 & 0.720\\
 & $ \frac{1}{2} \to \frac{1}{2} + 0$ & 1 & 0.953 & 0.918 & 0.916\\
 & & $\gamma_5$ & 1.057 & 1.132 & 1.179 \\ 
\hline
$1 \to 3 + \overline{3}$ & eikonal & --- & 0.802 & 0.695 & 0.659\\ 
 & eikonal $+ x_3^2$ & --- & 1.201 & 1.518 & 1.670 \\ 
\hline
$3 \to 3 + 1$ & $\frac{1}{2} \to \frac{1}{2} + 1$ & 1 & 
0.323 & 0.306 & 0.287 \\
 & & $\gamma_5$ & 0.356 & 0.365 & 0.349 \\
 & $\frac{1}{2} \to \frac{1}{2} + 0$ & 1 & 0.312 & 0.284 & 0.258 \\
 & & $\gamma_5$ & 0.357 & 0.363 & 0.344 \\
 & $0 \to 0 + 1$ & 1 & 0.287 & 0.242 & 0.218 \\
 & $0 \to 0 + 0$ & 1 & 0.279 & 0.224 & 0.194 \\
 & $0 \to \frac{1}{2} + \frac{1}{2}$ & 1 & 0.359 & 0.379 & 0.375 \\
 & & $\gamma_5$ & 0.347 & 0.354 & 0.346 \\
 & $\frac{1}{2} \to 0 + \frac{1}{2}$ & 1 & 0.294 & 0.257 & 0.239 \\
 & & $\gamma_5$ & 0.314 & 0.302 & 0.298 \\
\hline
$3 \to 3 + 8$ & $0 \to \frac{1}{2} + \frac{1}{2}$ & 1 & 
1.634 & 1.833 & 1.922 \\
 & & $\gamma_5$ & 1.574 & 1.712 & 1.775 \\
 & $\frac{1}{2} \to 0 + \frac{1}{2}$ & 1 & 
1.385 & 1.320 & 1.291 \\
 & & $\gamma_5$ & 1.549 & 1.664 & 1.675 \\
\hline
$8 \to 3 + \overline{3}$ & $\frac{1}{2} \to \frac{1}{2} + 0$ & 1 &
 0.561 & 0.493 & 0.445 \\
 & & $\gamma_5$ & 0.621 & 0.607 & 0.574 \\
\hline
\end{tabular}
\caption{Three measures on the amount of gluon radiation in different
processes, for mass ratios $r_1 = r_2 = 0.2$. For clarity, results have
been normalized to the process in the top line. See the text for a
detailed explanation.}
\label{tab:threejetrate}
\end{center}
\end{table}

In order to quantify the difference in the total amount of gluon 
radiation, we compare three measures on the three-jet phase space.
Since the total three-jet cross section contains a soft-gluon 
divergence, this soft region has to be avoided. One measure is thus to
integrate the total amount of radiated gluon energy, in shorthand
\begin{equation}
\int x_3 = \int x_3 \, \frac{1}{\sigma_0} \, 
\frac{\d \sigma}{\d x_1 \, \d x_2} \, \d x_1 \, \d x_2
~.
\end{equation} 
Correspondingly we define a shorthand $\int (1-x_1)(1-x_2)$ as an 
alternative removal of the denominator of the matrix elements 
(generalized to $\int (1+r_1^2-r_2^2-x_1)(1+r_2^2-r_1^2-x_2)$ for 
unequal masses). Finally a Durham distance \cite{Durham} 
$y_{ij} = \min(x_i^2,x_j^2) (1 - \cos\theta_{ij})/2$ is used to
define a hard three-jet region for which 
$y_{\mathrm{D}} = \min(y_{12},y_{13},y_{23}) > 0.1$
and a corresponding three-jet rate $\int \theta(y_{\mathrm{D}}-0.1)$,
where $\theta$ is the step function. To simplify a comparison between 
the processes, all results have arbitrarily been normalized to those 
for the $V \to \q\qbar$ process.

As can be seen in Table~\ref{tab:threejetrate}, the three measures 
give about the same message, namely that differences between the
processes are significant. In $\int x_3$ the ratio between the two 
extremes is a factor 1.30, in $\int (1-x_1)(1-x_2)$ 1.70, and in 
$\int \theta(y_{\mathrm{D}}-0.1)$ 1.91. This shows a steady 
progression of larger ratios the more one is biased towards the hard 
three-jet region. The eikonal expression is below all the calculated
processes, and the modified eikonal (full blast) above.

\begin{table}[t]
\begin{center}
\begin{tabular}{|c|c|c|c|@{\protect\rule[-2mm]{0mm}{7mm}}}
\hline
colour & spin & $\int (1-x_1)(1-x_2)$ & 
$\int \theta(y_{\mathrm{D}}-0.1)$ \\
\hline
$1 \to 3 + \overline{3}$ & $1 \to \frac{1}{2} + \frac{1}{2}$ 
& 1.000 & 1.000 \\
 & $0 \to \frac{1}{2} + \frac{1}{2}$  & 1.167 & 1.184 \\
 & $1 \to 0 + 0$  & 1.000 & 1.141\\
 & $0 \to 0 + 0$  & 0.667 & 0.773\\
 & $ \frac{1}{2} \to \frac{1}{2} + 0$ & 0.917 & 0.979\\
\hline
$1 \to 3 + \overline{3}$ & eikonal & 0.667 & 0.773\\ 
 & eikonal $+ x_3^2$ & 1.667 & 1.595 \\ 
\hline
$3 \to 3 + 1$ & $\frac{1}{2} \to \frac{1}{2} + 1$  & 0.347 & 0.282 \\
 & $\frac{1}{2} \to \frac{1}{2} + 0$  & 0.347 & 0.282 \\
 & $0 \to 0 + 1$  & 0.222 & 0.214 \\
 & $0 \to 0 + 0$  & 0.222 & 0.214 \\
 & $0 \to \frac{1}{2} + \frac{1}{2}$ & 0.389 & 0.328 \\
 & $\frac{1}{2} \to 0 + \frac{1}{2}$ & 0.264 & 0.260 \\
\hline
$3 \to 3 + 8$ & $0 \to \frac{1}{2} + \frac{1}{2}$ & 1.701 & 1.659 \\
 & $\frac{1}{2} \to 0 + \frac{1}{2}$ & 1.389 & 1.384 \\
\hline
$8 \to 3 + \overline{3}$ & $\frac{1}{2} \to \frac{1}{2} + 0$ 
& 0.573 & 0.487 \\
\hline
\end{tabular}
\caption{Two measures on the amount of gluon radiation in different
processes, for mass ratios $r_1 = r_2 = 0$, cf. 
Table~\protect\ref{tab:threejetrate}. For massless daughters the
$\gamma_5$ factor makes no difference, and so those results are not 
shown separately. The $\int x_3$ measure now is collinear
divergent and therefore not shown.}
\label{tab:threejetratezero}
\end{center}
\end{table}

The results in Table~\ref{tab:threejetrate} are for the case where
the daughter masses constitute a significant fraction of the energy
available. The other limit, where instead the daughters are massless,
is shown in Table~\ref{tab:threejetratezero}. The main message is that
the process dependence remains also in the latter case, even if
normally
reduced in magnitude by about a factor of two between the extremes.
The detailed picture is not so simple, however. Some processes agree 
in the massless limit when they do not for nonvanishing masses while, 
in the other extreme, others disagree even more for vanishing masses.

The set of  processes with colour structure $3 \to 3 + 1$ can be
viewed as crossed versions of the $1 \to 3 + \overline{3}$ ones,
but the differences in kinematics result in another overall
picture. Examples of radiation patterns are shown by dashed 
curves in Fig.~\ref{fig:compproc}. The small-angle behaviour displays
the universal dead cone effect at small $x_3$ and again diverges
wildly at larger $x_3$. Note that we have oriented all processes
so that the radiating daughter colour charge is at $0^{\circ}$,
wherefore the radiation continues to drop off at large angles.
This is unlike the previous set of curves, which turn around at or 
near the mid angle to the other radiating daughter. (For small
$x_3$ the two daughters are almost back-to-back, i.e. a bisector
at $90^{\circ}$, while an $x_3$ of 0.6 allows an almost symmetric
configuration with $120^{\circ}$ between all three.) 

\begin{figure}[tp]
\begin{center}
\epsfig{file=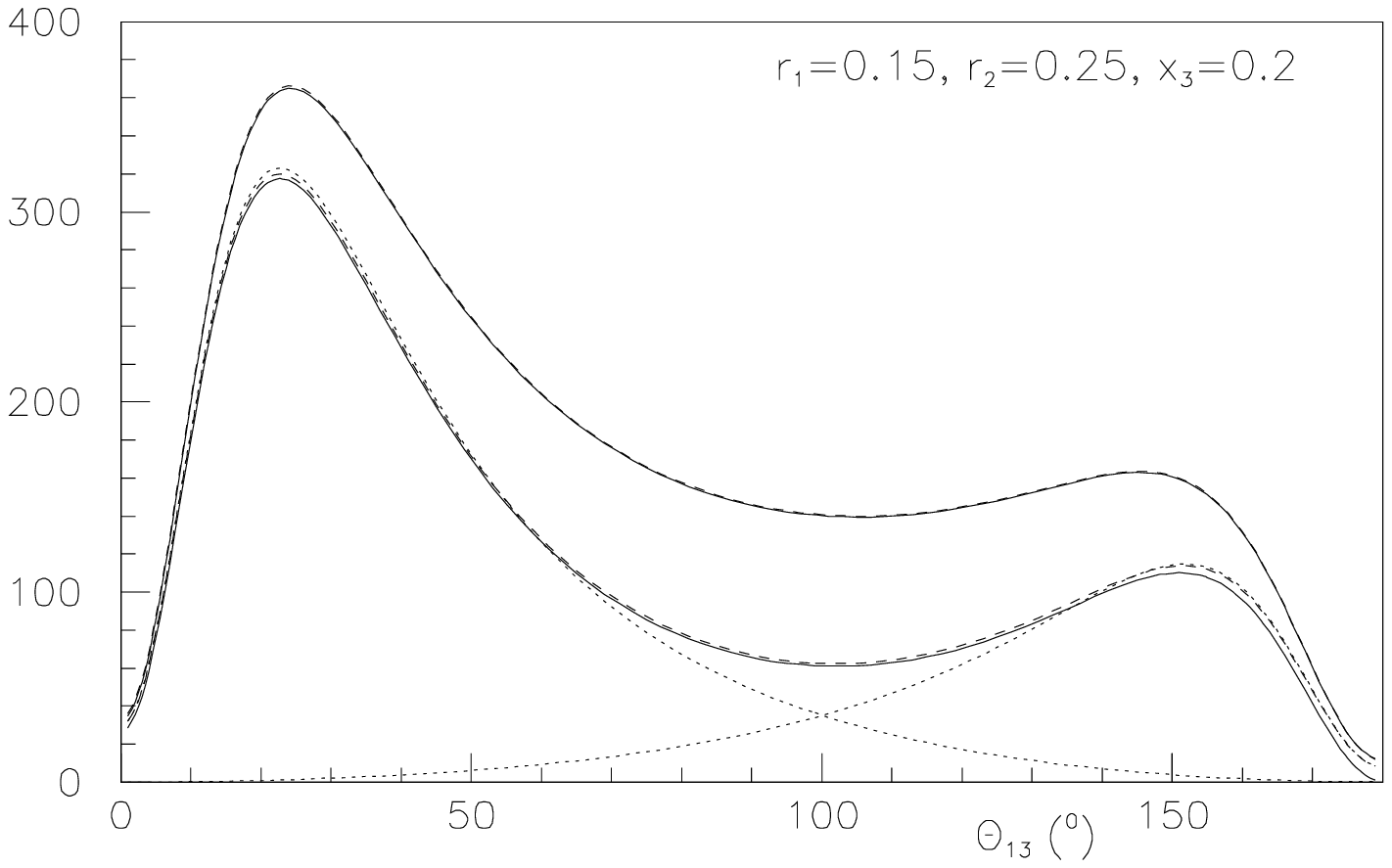}
\end{center}
\vspace{-2\baselineskip}
\caption{Test of the additivity of gluon emission rates, 
$\d \sigma / \d x_1 \, \d x_2$ with a normalization factor
$\sigma_0 \, C_F \, \alphas/2\pi$ removed. The dotted curves
show two processes with colour flow $3 \to 3 + 1$ and
$3 \to 1 + 3$, respectively (and spin $1/2 \to 1/2 + 1$). 
The upper and lower full curves give the complete expressions for 
$1 \to 3 + \overline{3}$ (spin $1 \to 1/2 + 1/2$) and  
$8 \to 3 + \overline{3}$ (spin $1/2 \to 1/2 + 0$) processes, 
respectively. The two dashed curves, almost completely hidden 
by the full ones, are the same processes according to the additive 
approximations in eq.~(\protect\ref{addradiation}).} 
\label{fig:addproc}
\end{figure}

It is here interesting to remind of the QED answer for the process
$V \to \f\fbar\gamma$ with $m_{\f} = m_{\fbar}=0$ \cite{grose}
\begin{equation}
\frac{1}{\sigma_0} \, \frac{\d \sigma}{\d x_1 \, \d x_2}
 = \frac{\alphaem}{2\pi} \frac{x_1^2 + x_2^2}{(1-x_1)(1-x_2)} 
 \left( e_{\f} \frac{1-x_1}{x_3} - e_{\fbar} \frac{1-x_2}{x_3}
 \right)^2 ~,
\label{QEDradiation}
\end{equation}   
where $e_V = e_{\f} + e_{\fbar}$ has been used to eliminate the
explicit appearance of terms corresponding to radiation off the $V$. 
For $e_V = 0$, $e_{\f} = - e_{\fbar}$ this is the QED analogue of 
eq.~(\ref{MEmassless}), while the current case corresponds to 
$e_{\fbar} =0$, $\alphaem e_{\f}^2 \to \alphas C_F$:
\begin{equation}
\frac{1}{\sigma_0} \, \frac{\d \sigma}{\d x_1 \, \d x_2}
 = \frac{\alphas}{2\pi} \, C_F \, \frac{x_1^2 + x_2^2}{(1-x_1)(1-x_2)} 
 \left( \frac{1-x_1}{x_3} \right)^2 
 = \frac{\alphas}{2\pi} \, C_F \,  \frac{x_1^2 + x_2^2}{x_3 (1-x_2)} 
 \, \frac{1-x_1}{x_3} ~.
\end{equation}   
Here the first part of the final expression essentially is the shower 
answer of eq.~(\ref{PSfirsthist}). In the soft-gluon limit
$(1-x_1)/x_3 \approx (1+\cos\theta_{13})/2 \approx (1-\cos\theta_{23})/2$,
so this extra factor gives a further dampening at large emission angles, 
above the shower ansatz (which in itself is dampened by one such angular 
factor relative to the colour singlet decay). A main consequence of this
extra factor is that the three-jet activity in a colour $3 \to 3 + 1$ 
event is less than half of that of a $1 \to 3 + \overline{3}$ one,
a pattern which remains when masses are included, see 
Table~\ref{tab:threejetrate}.   

Finally we come to the processes with a gluino, denoted by 
dash-dotted lines in Fig.~\ref{fig:compproc}. In the limit of
infinitely many colours $N_C$, the process $3 \to 3 + 8$ may be viewed
as one colour flowing through from the initial triplet to the
gluino, and a different colour-anticolour pair created between
the final triplet and the gluino, i.e. as the sum of
$3 \to 1 + 3$ and $1\to 3 + \overline{3}$. Interference terms
between the two colour flows would be suppressed by a factor
$1/N_C^2$, just as in the radiation pattern for the colour-related
process $V \to \q\qbar\g$ \cite{Leningraddipole}. This process 
therefore shows the most three-jet activity of the ones studied, 
especially in the not-displayed gluino hemisphere. 

By the same token, $8 \to 3 + \overline{3}$ may be approximated by 
the incoherent sum of a $3 \to 3 + 1$ and a 
$\overline{3} \to 1 + \overline{3}$ radiation pattern, which gives less 
radiation than the $1 \to 3 + \overline{3}$ processes. The QED formula,
eq.~(\ref{QEDradiation}), may here offer a convenient starting point.
The $1 \to 3 + \overline{3}$ process corresponds to 
$e_{\f} = - e_{\fbar}$, so that the interference term 
$-e_{\f}e_{\fbar}$ is positive. The $8 \to 3 + \overline{3}$
process, e.g. $\sg \to \q\sqbar$,  can instead be emulated in QED
by $e_{\f} =  e_{\fbar}$, i.e. a doubly-charged gluino 
\cite{QEDanalogue}. The interference term then is negative, although 
suppressed by a colour factor $1/N_C^2$ in the QCD case. 

Thus, given a radiation pattern $f_1(\theta_{13})$ for emission
off parton 1 like in a colour $3 \to 3 + 1$ process, and a corresponding
$f_2(\theta_{13})$ for a colour $3 \to 1 + 3$ one (essentially
obtainable by swapping the kinematics of the above process,
$x_1 \leftrightarrow x_2$, $r_1 \leftrightarrow r_2$) one may
guess at the radiation pattern for the  $1 \to 3 + \overline{3}$ 
and $8 \to 3 + \overline{3}$ processes: 
\begin{eqnarray}
f_{1 \to 3 + \overline{3}}(\theta_{13}) & = & f_1(\theta_{13}) +
f_2(\theta_{13}) + 2 \, \sqrt{f_1(\theta_{13})f_2(\theta_{13})}
~, \nonumber \\
f_{8 \to 3 + \overline{3}}(\theta_{13}) & = & f_1(\theta_{13}) +
f_2(\theta_{13}) - \frac{2}{9} \, 
\sqrt{f_1(\theta_{13})f_2(\theta_{13})} ~.
\label{addradiation}
\end{eqnarray}
This turns out to be a good approximation up to fairly large $x_3$ 
values, i.e. so long as the spin structure is not too important, 
see e.g. Fig.~\ref{fig:addproc}. The importance of the interference 
term also gives a simple explanation for the difference between the 
height of the peaks in the full and dashed curves in 
Fig.~\ref{fig:compproc}a--b.

\subsection{Parity dependence}
\label{sec:misc}

\begin{figure}
\begin{center}
\epsfig{file=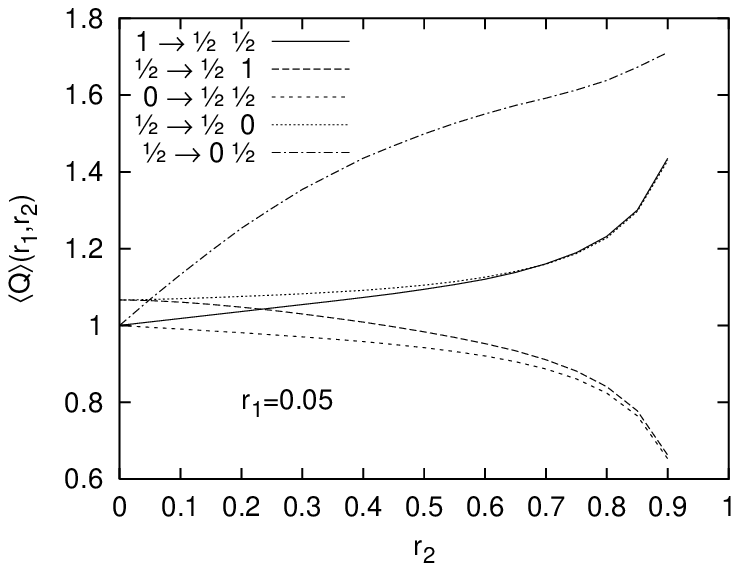}
\vspace{\fill}
\epsfig{file=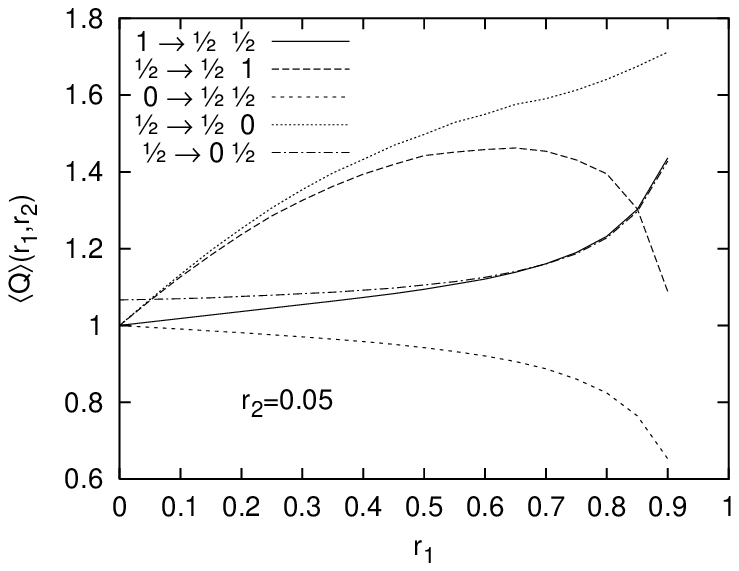}
\epsfig{file=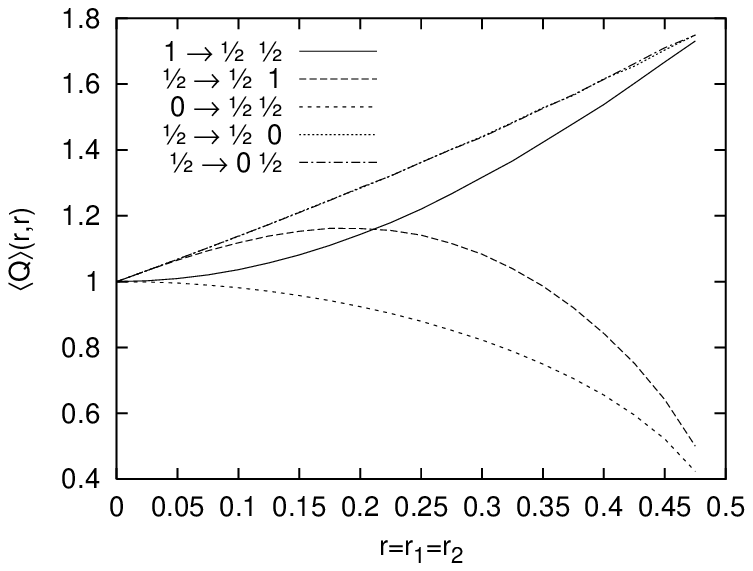}
\end{center}
\caption{$\langle Q \rangle (r_1,r_2)$, eq.~(\ref{eq:Qdef}),
with $r_1$ fixed, $r_2$ fixed, or $r=r_1=r_2$. The processes are grouped
according to spin structure.
}\label{Qratio}
\end{figure}

Let us further quantify differences induced by having or not a $\gamma_5$ 
factor in the matrix element, i.e. between vector ($V$) and axial vector
($A$) sources, between scalar ($S$) and pseudoscalar ($P$) ones, etc.
We will use a measure which is the mean of the matrix element ratios
over the whole phase space for gluon emission. Since the matrix elements
with and without the $\gamma_5$ coupling both have the same divergence
structures the ratio is well-behaved everywhere. All phase space points
are here given equal weight, so large differences may not always be
reflected in significant changes of the radiation pattern.
To be specific, we study the function
\begin{equation}
\langle Q \rangle (r_1,r_2) =
\frac{1}{\int \d x_1 \d x_2} \int Q(x_1,x_2,r_1,r_2) \d x_1\d x_2,
\label{eq:Qdef}
\end{equation}
where
\begin{equation}
Q(x_1,x_2,r_1,r_2) =
\frac{\sigma^1_0}{\sigma^{\gamma_5}_0}
\frac{\d \sigma^{\gamma_5}/\d x_1 \d x_2}
     {\d \sigma^1/\d x_1 \d x_2}.
\end{equation}
$Q(x_1,x_2,0,0)$ is equal to unity in all of phase space and in most
cases $Q(x_1,x_2,r,0)$ is also equal to unity. The exception is when the
$r=0$ mass is a boson (spin 0 or 1) and the $r \neq 0$ one is spin \half. 
All the processes in Table~\ref{tab:processes} which have both
non-$\gamma_5$ and $\gamma_5$ couplings are represented in
Fig.\ref{Qratio}. Processes with the same spin structure have the same
ratio, where both the spin of the initial and final states, as well as
the order of the decay products, are of importance. We notice that the
difference between e.g. vector and axial vector couplings can be rather
large in some cases, even for small and intermediate masses. The most
important case, however, with two fermions in the final state, show
small differences for $r<0.5$, so this aspect will only be significant
in the case of top production.  Physical consequences will be
investigated in Section~\ref{sec:applications}.

\section{Applications}
\label{sec:applications}
In this section, the matrix element corrected shower is used to study
some processes at current and future colliders.
It is not to be seen as a comprehensive review, but as simple examples
intended to illustrate the main features.

\subsection{Bottom in $\Z^0$ decay}
\label{sec:bottom}

We start by examining a process where data with good statistics already
exists and detailed comparisons are possible, namely gluon radiation off
bottom quarks produced in the process $\ee \to \Z^0 \to \b\bbar$ at the
$\Z^0$ pole. Gluons are not observed directly in the final state, instead 
they materialize as jets of hadrons. We will mainly consider jets on the
parton level, but we also study the effects of hadronization and decays.
The experimental data with which our results are compared have already
been corrected to the parton level, using, among others, the same model
for fragmentation which we will use.

Jets are constructed by considering all pairs of particles $(i,j)$ in an
event, where the particles can be hadrons, partons or clusters of hadrons 
or partons, finding the pair with the smallest ``distance'' $y_{ij}$.
If this number is smaller than the ``jet resolution'' parameter, $y_c$,
the pair is joined into a new cluster by summing up their four-momenta.
This procedure is repeated until all $y_{ij}$ are larger than $y_c$.
The resulting clusters are the jets in the event at the resolution scale
$y_c$. Jet algorithms differ mainly by the definition of $y_{ij}$, and
several jet measures have been proposed in the literature
\cite{Durham,jet-measures}. Following the lead of the LEP analyses we
want to compare with, we have settled on the {\sc Durham} \cite{Durham}
algorithm which defines
\begin{equation}
y_{ij} = \frac{2\mathrm{min}(E_i^2,E_j^2)(1-\cos\theta_{ij})}
{E^2_\mathrm{vis}}.
\end{equation} 
For small angles, this is approximately the relative transverse momentum
squared of the pair, scaled to the visible energy in the event. If jets
are constructed with a large $y_c$, only the most energetic partons will
be resolved, and most events will become 2-jet events. When $y_c$ is
decreased, smaller structures will start to emerge and multi-parton final 
states become more important. It is in this region that the parton shower 
approach to gluon radiation is most useful.

For a primary quark flavour q, the $n$-jet rate is defined as
\begin{equation}
R_n^q(y_c) = \frac{\sigma_{\q\qbar \to n~\mathrm{jets}}(y_c)}
{\sigma_{\q\qbar \to \mathrm{hadrons}}},
\end{equation}
which for $n > 2$ is generally a decreasing function of $y_c$. We want
to study the difference between gluon radiation off b quarks and off
light quarks. A suitable observable is then the ratio between the
respective n-jet rates
\begin{equation}
R_n^{bl}(y_c) = \frac{R_n^b(y_c)}{R_n^l(y_c)},~\mathrm{with}~l=\mathrm{u,~d~or~s.}
\end{equation}

Experiments at LEP \cite{R3bl-Delphi,R3bl-Aleph} have found both
$R_3^{bl}$ and $R_4^{bl}$ to be smaller than one, approaching unity for
large $y_c$. This can be understood qualitatively as a consequence of
the well-known dead cone effect \cite{deadcone}, stating that the
radiation of collinear gluons off heavy quarks is suppressed. The
production of well separated jets, however, is not significantly
suppressed and should approach that of the light quarks for large $y_c$.

We want to study this effect more quantitatively by using our improved,
matrix element corrected, parton shower. The new approach, described in
Section~\ref{sec:new}, has been implemented in \Py~6.153. This is
compared to \Py~6.152 containing the older approach of
Section~\ref{sec:older}, which includes the correct massive matrix
element correction in the first emission only. As a reference, we also
include results from the algorithm implemented in \Py~6.129, where
mass-effects are incorrectly included in the matrix element correction
of the first emission, cf. Section~\ref{sec:old}.

\begin{figure}
\begin{center}
\epsfig{file=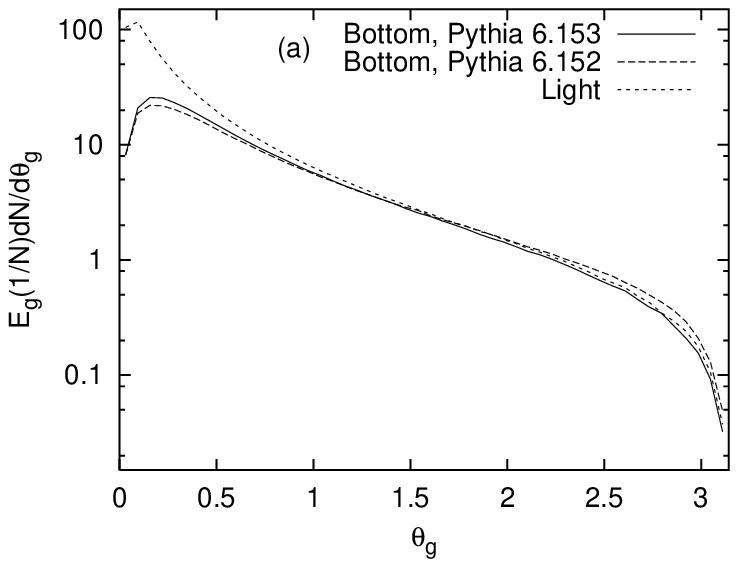}
\vspace{\fill}
\epsfig{file=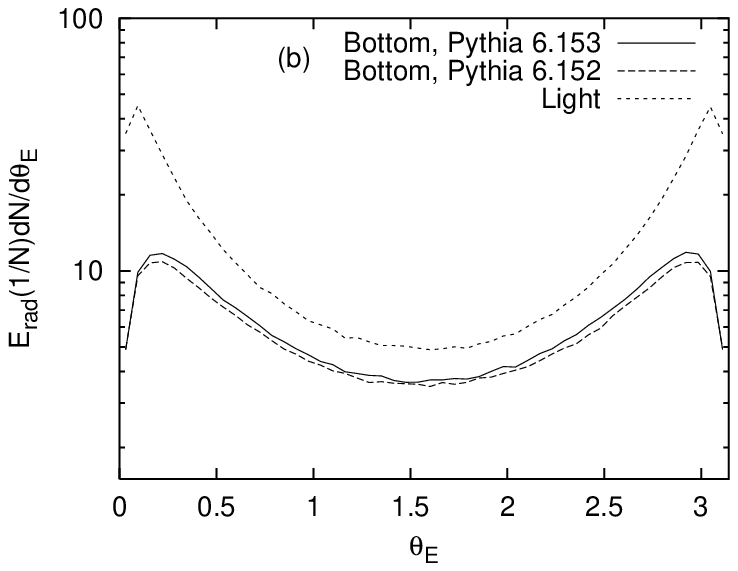}\\
\epsfig{file=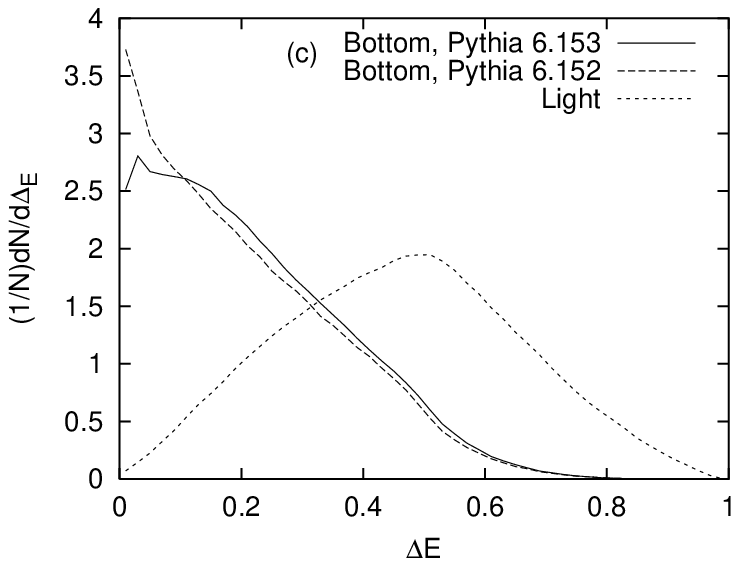}
\vspace{\fill}
\epsfig{file=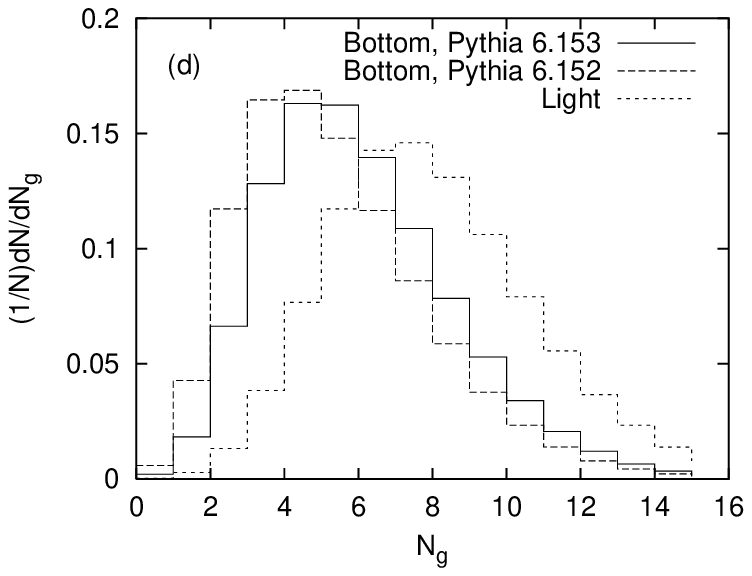}
\end{center}
\caption{Gluon radiation and energy flow in the process
$\ee \to \Z^0 \to \q\qbar$, where $\q$ is either a light (u, d, s) quark
or a b quark. The b-mass is here set to the default in \Py, 4.8 GeV.
(a) Energy weighted gluon angle distribution.
(b) Energy flow.
(c) Radiated energy fraction, $\Delta E =  E_\mathrm{rad}/E_\mathrm{CM}$.
(d) Gluon multiplicity.
}\label{fig:energyflow}
\end{figure}

The jet rate reflects both the amount of energy radiated and the
direction in which it is radiated. Fig.~\ref{fig:energyflow}a shows the
gluon radiation pattern in the process $\ee \to \Z^0 \to \q\qbar$ at
the $\Z^0$-pole, for the old and the new shower routines both for light
(u, d and s) quarks and heavy (bottom) quarks. The angle, $\theta_g$,
is defined as the angle between the radiated gluon and the primary quark
in the CM system of the primary quark pair. In principle, gluons are
radiated by the $\q\qbar$ dipole as a whole, but in our implementation
each radiated gluon is assigned to an initial quark, and in the collinear 
limit this separation is quite sensible. In the case of the total energy
flow, Fig.~\ref{fig:energyflow}b, we add the two contributions, and
include the energy taken by quarks produced in gluon splitting, so two
symmetrical peaks appear. Fig.~\ref{fig:energyflow}c shows the
distribution of $\Delta E = E_\mathrm{rad}/E_\mathrm{CM}$, i.e. the
total energy fraction radiated in the shower.
Fig.~\ref{fig:energyflow}d, finally, shows the gluon multiplicity in the
shower, which obviously is quite dependent on the lower shower
cut-off $Q_0$.

We note first of all that the difference between light and heavy quarks
is largest in the region of small angles, the dead cone, and also that
the radiated energy fraction is much larger for light quarks, peaked
around 0.5 if $Q_0=1$ GeV is used to remove the collinear emission off
light quarks. For heavy quarks, this collinear emission is regulated by
the mass. For large angles, the light and heavy quark distributions
converge as they should. In the new approach, the amount of gluon
radiation in b events at small angles is somewhat enhanced relative to
the old one, where the dead cone was slightly exaggerated, but
differences are small. By the change of evolution variable,
eq.~(\ref{eq:newQ2}), the fraction of events without any shower at all
($\Delta E=0$) has decreased by almost a factor of 3 from about 4 per
mille to 1.4. As a consequence, the peak at $\Delta E=0$ has been
flattened out. The gluon multiplicity has also increased slightly.

We now want to investigate how the changes in the algorithm affect
$R_3^{bl}$ and $R_4^{bl}$. Since one is studying small deviations from
unity, these measures are very sensitive to changes in the gluon
radiation pattern. Also, the LEP experiments have large samples of
$\Z^0$ events and studies at the per cent level are feasible. We will
mostly study the behaviour of the model on the parton level, comparing
different alternatives. Here the parton level is defined by the partonic
configuration at the shower cut-off scale $Q_0$, below which no further
emissions occur. But first we should comment on the effects of
fragmentation. Below the $Q_0$ scale, the Lund string fragmentation model 
\cite{Lundstring} describes how the partons transform into the primary
hadrons. Subsequently these may decay further. Once a $Q_0$ has been
chosen, the parameters of the fragmentation model should be fitted to
data. Here we will only study the variation with $Q_0$, both at the
parton level and the hadron level, without any such retuning. At the
hadron level, we consider first the primary hadrons and study decay
effects separately.

\begin{figure}
\begin{center}
\epsfig{file=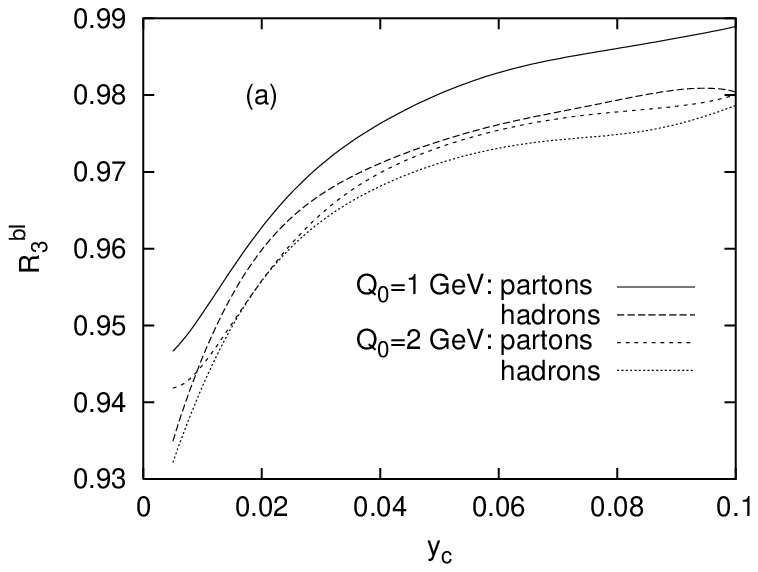}
\vspace{\fill}
\epsfig{file=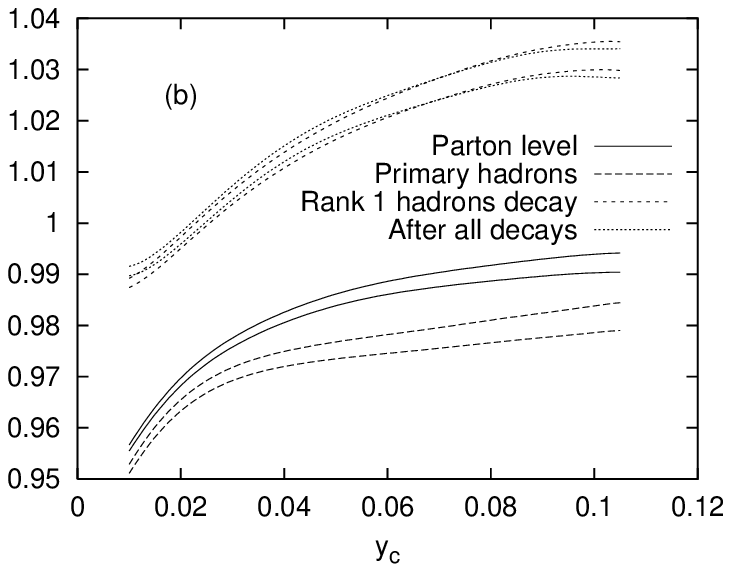}
\end{center}
\caption{Study of fragmentation and decay effects. (a) Effects of
changing the cut-off, $Q_0$, in the shower, both on the parton level and
at the level of primary hadrons (i.e. before decays). (b) Effects of
hadronization, decay of rank 1 hadrons and decay of all hadrons. In (b)
each alternative is represented by the $\pm~1\sigma$ curves given by the
Monte Carlo statistics.
}\label{fig:frag}
\end{figure}

Fig.~\ref{fig:frag}a shows the effect of fragmentation and the $Q_0$
dependence. We calculate $R_3^{bl}(y_c)$ at the parton level and at the
level of primary hadrons for $Q_0=1$ and 2 GeV, respectively. The $Q_0$
dependence is largest on the parton level with $R_3^{bl}$ slightly lower
for the larger $Q_0$. This is because the increase in $Q_0$ increases
the 3-jet rate for light quark events, while heavy-quark events are not
influenced as much: collinear emissions, which dominate at the end of the 
cascade, are suppressed anyway. Intuitively one might have expected the
jet rate to be higher for a lower $Q_0$, since this corresponds to a
larger number of partons, but the further partons emitted between 1 and
2 GeV cannot give rise to new jets of their own. They only smear out the
energy of the existing jets and possibly make them fall below the cut-off.
Furthermore, the hadronization stage tends to decrease the $Q_0$
dependence because the Lund model is infrared safe and collinear gluon
emissions do not affect the string fragmentation process. A retuning of
the fragmentation parameters would further limit the effect of small
changes in $Q_0$.

Fig.~\ref{fig:frag}b shows the effects of fragmentation and decay.
The rank 1 hadrons are the ones that contain the primary quarks from the
decay of the $\Z^0$. In the case of bottom production, the rank 1 hadrons 
will be mainly B mesons and $\Lambda_\b$ baryons. Because of the hard
fragmentation function for heavy hadrons, the multiplicity of primary
hadrons in b events is smaller than that in a light quark event,
reflected in the lower $R_3^{bl}$ curve. However, once the heavy hadrons
have decayed, their decay products will more than compensate this. Thus,
if the primary hadrons are allowed to decay, the value of $R_3^{bl}$
lies above unity for $y_c>0.03$. So, on the one hand, heavy hadrons take
a large part of the primary quark energy, but they also decay to many
particles. For $R_3^{bl}$ there is not a large difference between
allowing all primary hadrons to decay or only the rank 1 ones, which can
be seen in the small difference between the two top curves in
Fig.~\ref{fig:frag}b. In the following, we will compare our results to
experimental data on the parton level, bearing in mind that this
comparison can be ambiguous in view of the dependence on $Q_0$. Our aim
here is not to achieve a perfect fit to data, but merely ensure that the
results are at the right level. As we will see, the data is anyway not
good enough to make precise discriminations between all different model
variations.
\begin{figure}
\begin{center}
\epsfig{file=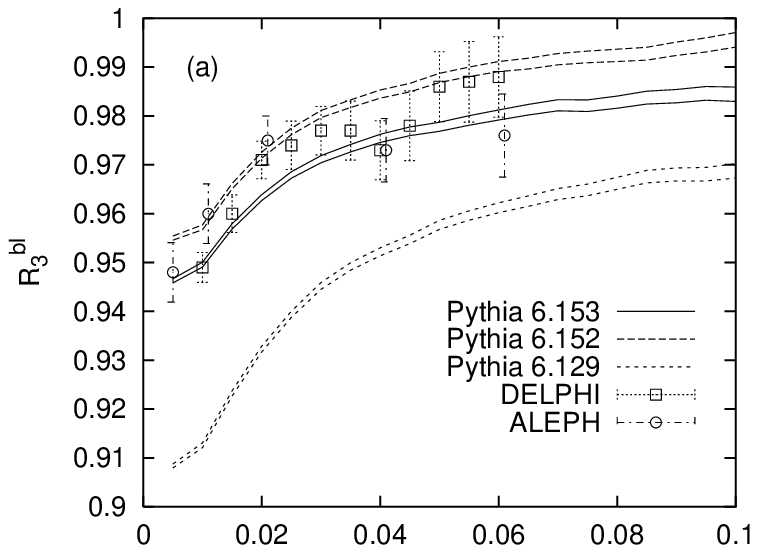}
\epsfig{file=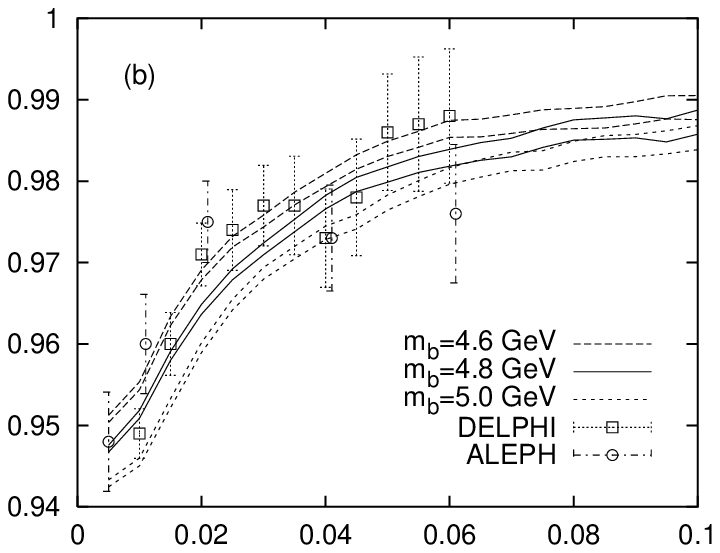}
\epsfig{file=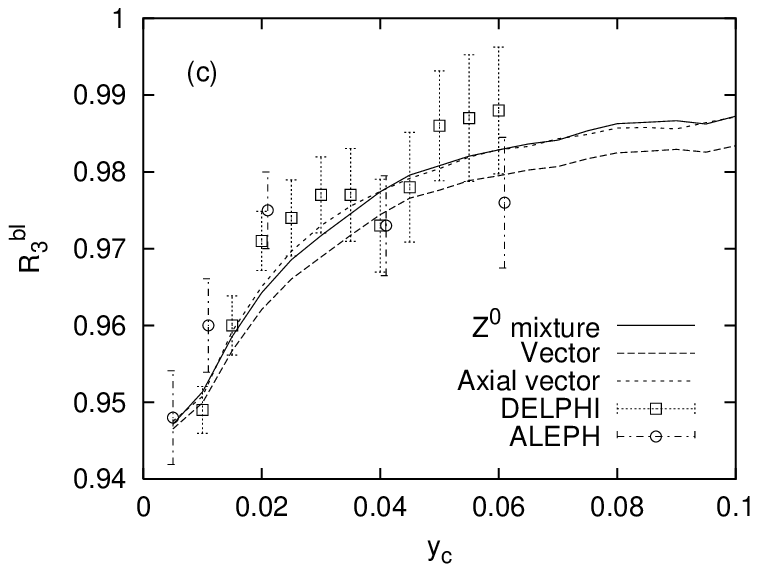}
\epsfig{file=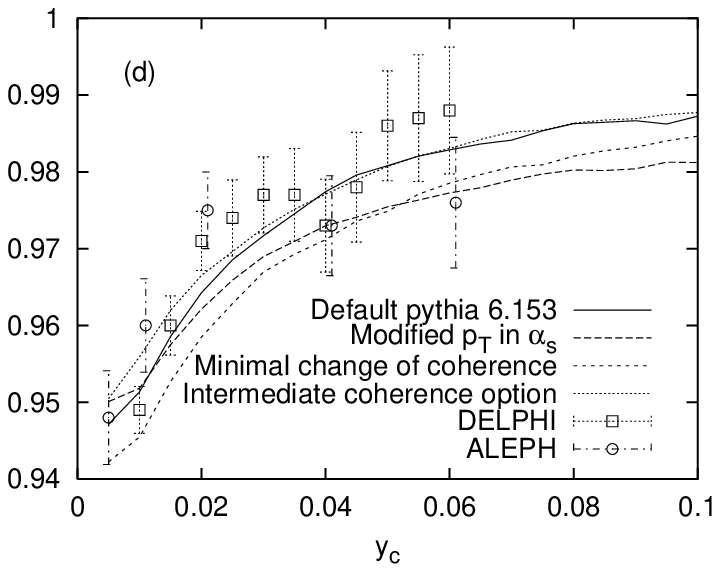}
\end{center}
\caption{$R_3^{bl}(y_c)$ for different model parameters and variations.
(a) Different main versions. Some variations of the latest version
(6.153): (b) Different bottom masses.
(c) Different sources.
(d) Other minor variations of the main theme, see the text for details.
The data are from \cite{R3bl-Delphi} and \cite{R3bl-Aleph}.
The data points from {\sc Aleph} (except the first) have been shifted
.001 units in $y_c$ for clarity. In (a) and (b) the $\pm 1\sigma$ curves
are given, while only the central value is shown in (c) and (d) for
clarity, but the statistics is comparable.
}\label{fig:R3bl}
\end{figure}
\begin{figure}
\begin{center}
\epsfig{file=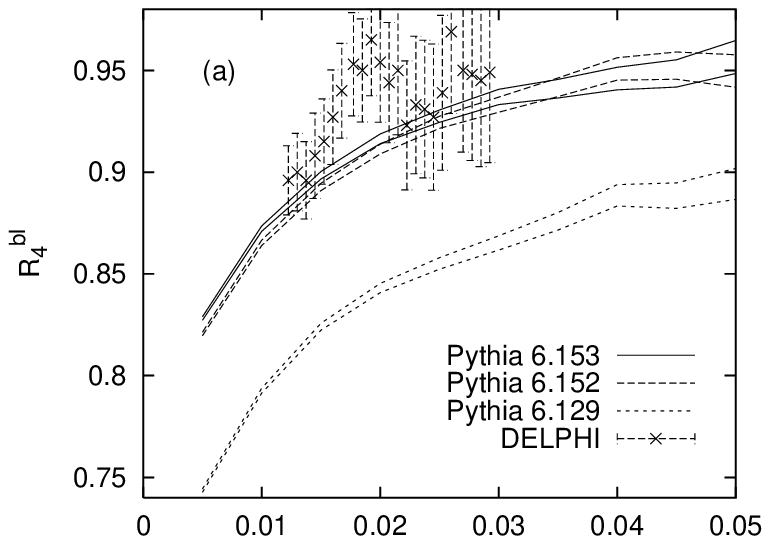}
\epsfig{file=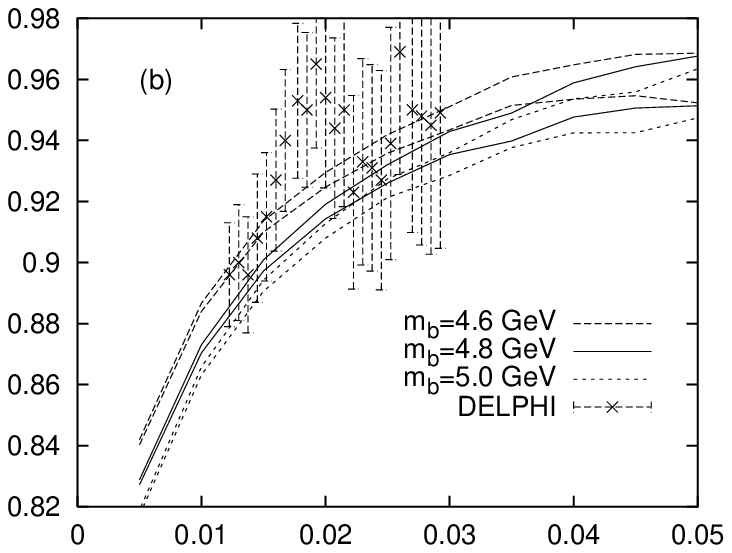}
\epsfig{file=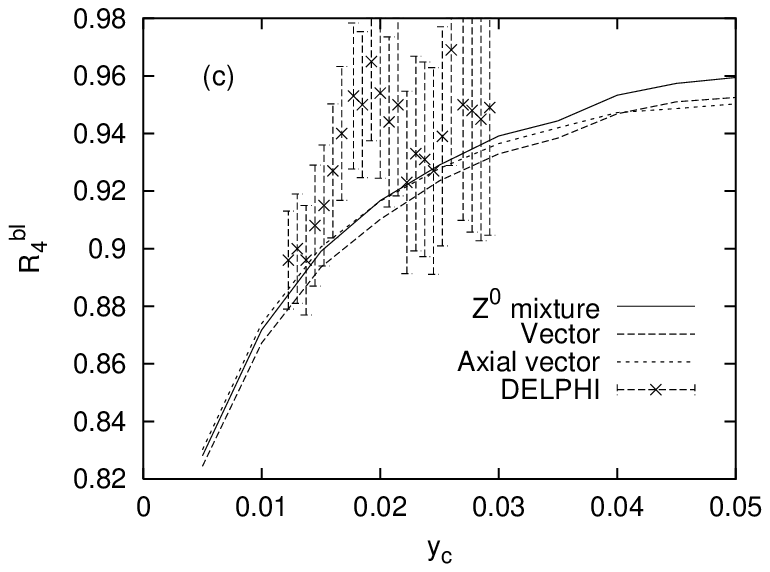}
\epsfig{file=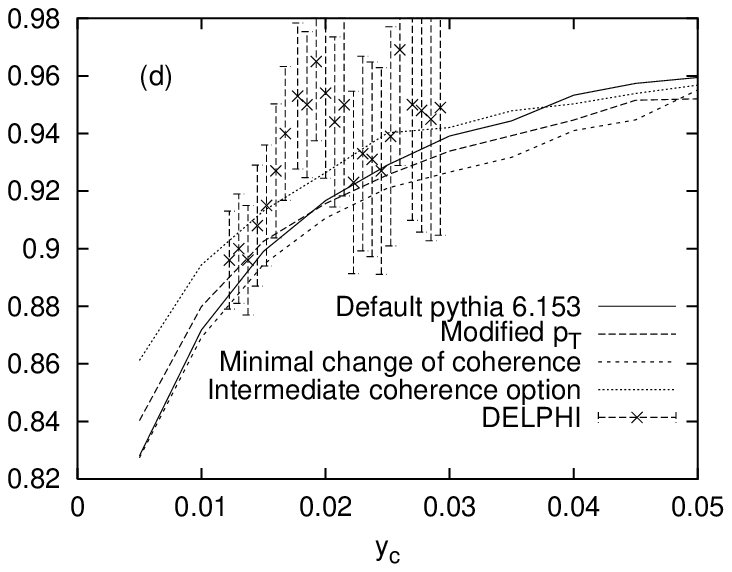}
\end{center}
\caption{$R_4^{bl}(y_c)$ for different model parameters and variations.
See caption Fig.~\ref{fig:R3bl} for details.
}\label{fig:R4bl}
\end{figure}

In Figs.~\ref{fig:R3bl}--\ref{fig:R4bl}, $R_3^{bl}$ and $R_4^{bl}$ are
shown as functions of $y_c$ for several model variations.
Figs.~\ref{fig:R3bl}a,~\ref{fig:R4bl}a show the difference between the
old and the modified shower models. An even older version, where mass
effects are exaggerated, is also shown as reference. Each model curve is
displayed as a one sigma band, showing the size of the Monte Carlo error
for $15 \cdot 10^{6}$ events of each kind. From the increase in
multiplicity and energy flow in b events in the new model, one would
naively expect the 3- and 4-jet rates to increase for heavy quarks, thus
increasing $R_n^{bl}$. Actually, the main effect is to reduce the 3-jet
rate for heavy quarks, again illustrating that allowing more
collinear/soft emissions will not necessarily increase the rate of well
separated jets. The naively expected increase of $R_4^{bl}$ (from
subsequent branchings) is balanced by a corresponding ``smearing'' loss
as for $R_3^{bl}$, giving only a small net effect.

As already noted, the old and new models differ mainly in the treatment
of subsequent branchings; both include matrix element corrections to the
first gluon emission on each side of the event. A clear discrimination
between these two versions is not possible, especially in view of the
variations that follow below and the relatively large experimental errors.
It is clear, however, that the algorithm in \Py~5.129 is ruled out by
the data.

Next we consider variations to the new default, in order to assess the
uncertainties of the model:
\begin{Itemize}
\item The bottom mass.
\item Parity dependence (vector vs axial vector source).
\item Change of $\pT^2$ argument in $\alphas$ from $z(1-z) m_a^2$ to
$z(1-z) m_a^2(1-m_b^2/m_a^2)^2$ in a branching $a \to bc$,
with $a$ being a heavy quark.
\item Coherence effects.
\end{Itemize}
The main free parameter is the bottom mass, and Figs.~\ref{fig:R3bl}b,
\ref{fig:R4bl}b show the result of varying this mass between 4.6 GeV and
5 GeV. Fits of NLO QCD calculations to LEP data give a value around 3 GeV 
for the running bottom mass in the $\overline{\mathrm{MS}}$
renormalization scheme \cite{QCDWG,R3bl-Delphi,R3bl-Aleph} at
renormalization scale $m_\Z$. Our model is based on LO matrix elements
and the mass in our case is the constituent quark mass, so they need not
agree. However, a somewhat lower mass than the default seems to be
favoured, especially for $R_4^{bl}$.

In sections \ref{sec:radpat} and \ref{sec:misc} we saw that there are
slight differences between different sources. The relevant mixture
for $\Z^0$ is given by eq.~(\ref{gamZmix}), but in Figs.~\ref{fig:R3bl}c, 
\ref{fig:R4bl}c we study the two extreme cases of a pure vector source
and a pure axial vector one. At the $\Z^0$-pole, the axial vector
component dominates, as can be seen in the figure, but at higher energies 
the vector one will take over. We see that the differences are
non-negligible, but not as large as the shown mass dependence, at least
for small $y_c$.

Finally, in Figs.~\ref{fig:R3bl}d, \ref{fig:R4bl}d, some further aspects
of the model, discussed in Section~\ref{sec:additional}, are varied.
Both the modified $\pT$ in $\alphas$ and the ``minimal'' change of
coherence tends to reduce $R_3^{bl}$, again showing how an increase in
the total amount of radiation need not give more separate jets. The
``intermediate'' coherence option, which still is a realistic alternative,
does give an increase, especially of $R_4^{bl}$. Thereby the overall
agreement with data is improved, as much as with a reduced $\b$ mass,
although uncertainties are sufficiently large that no firm conclusions
should be drawn.

Other minor issues could be studied and here we just mention a few.
The definition of $R_n^{bl}$ used here is based on a classification of
events with heavy or light {\it primary} quarks. An alternative is to
use a ratio between b-tagged and anti b-tagged events. The difference
here is in the classification of events with gluon splitting,
$\g \to \b\bbar$, in the shower. For the $R_3^{bl}$ ratio this is a
minor issue, but it is non-negligible in the case of $R_4^{bl}$
\cite{QCDWG}. This uncertainty is already included in the error-bars for
the experimental points in Fig.~\ref{fig:R4bl}, so need not be studied
separately here. Another issue is the energy dependence of the results.
Obviously the mass effects will decrease for larger energies, because of
the decrease of $r=m_\b/\sqrt{s}$, and this has already been studied at
189 GeV \cite{QCDWG}. At the same time, however, the fraction of vector
coupling increases significantly above the $\Z^0$-pole, thus decreasing
$R_3^{bl}$, cf. Fig.~\ref{fig:R3bl}c. At intermediate energies this could 
therefore give rise to a partial cancellation of effects.

\subsection{Bottom in Higgs decay}
\label{sec:Higgs}

We now turn to gluon radiation in Higgs decay. In the standard model,
the decay $\H^0 \to \b\bbar$ is expected to be large, or even dominate,
for Higgs masses up to the $\W^+\W^-$ threshold, and in extensions to the 
standard model also heavier Higgs states can have significant $\b\bbar$
branching ratios. When the fragmentation function for B mesons, i.e. the
distribution of $z=2E_\B/\sqrt{s}$, is measured at LEP1, the b quark is
produced from a spin 1 source. If instead the source is a Higgs boson,
the difference in gluon radiation could give rise to a changed
fragmentation function. Such a change would influence the experimental
vertex detection efficiency which, if uncorrected, would give rise to
errors in the determination of cross sections. It is therefore important
to be able to describe in detail the gluon radiation pattern in this
decay. A different measure of the nature of the source is the jet
topology, which is also studied in the form of a modified $R_3^{bl}$
ratio, cf. the preceding section.

\begin{figure}
\begin{center}
\epsfig{file=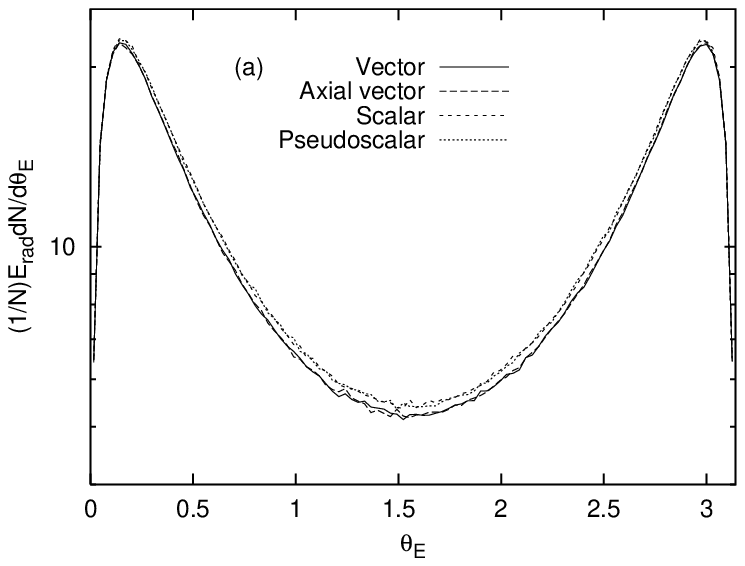}
\vspace{\fill}
\epsfig{file=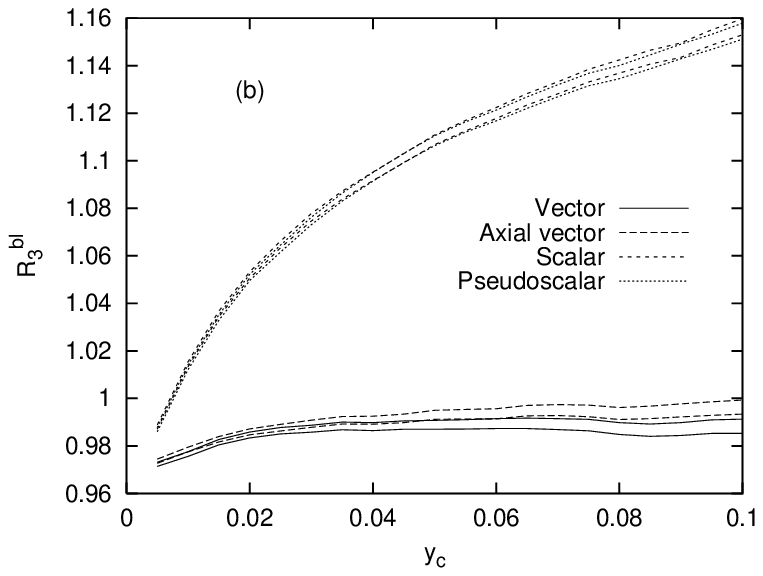}\\
\epsfig{file=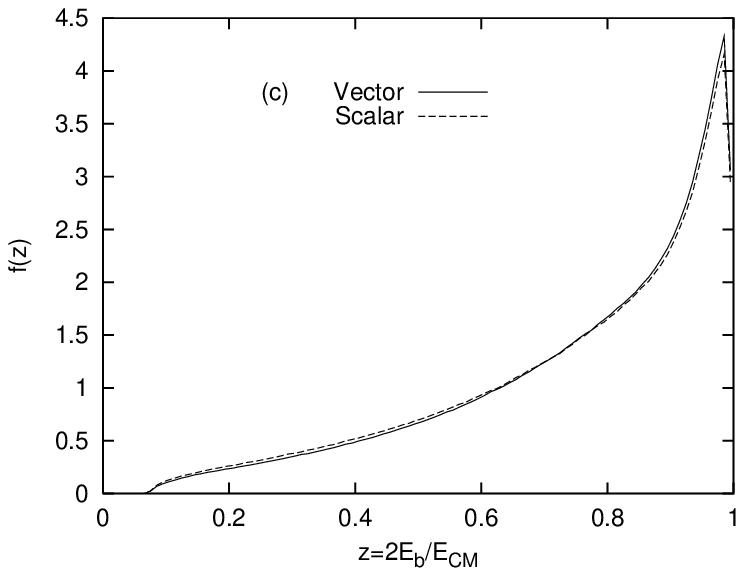}
\vspace{\fill}
\epsfig{file=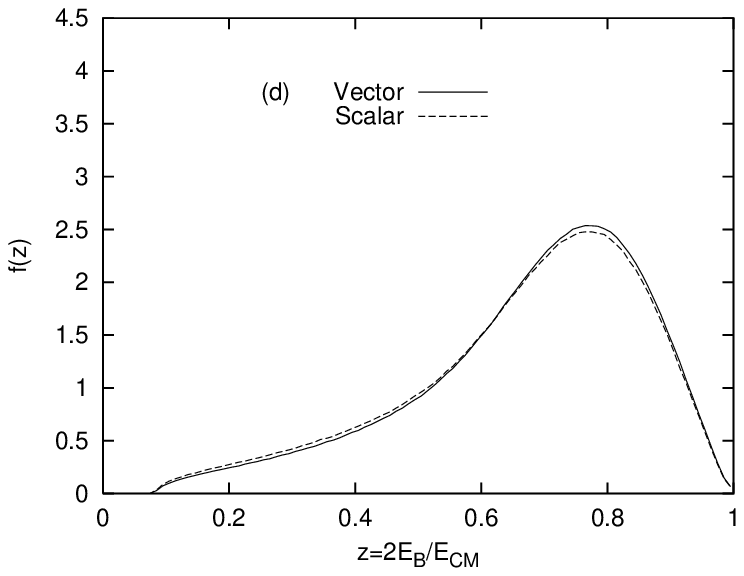}
\end{center}
\caption{Comparison between vector, axial vector, scalar and pseudoscalar 
sources of $\b\bbar$ pairs at 130 GeV.
(a) Energy flow in the shower.
(b) 3-jet rate, normalized to $\gamma^*/\Z^* \to \q\qbar$,
i.e. $R_3^{bl}(y_c)=R_3^b(X \to \b\bbar)/R_3^l(\gamma^*/\Z^* \to \q\qbar)$,
where $X$ is $V$, $A$, $S$ or $P$ and q is a light flavour.
$1\sigma$ Monte Carlo error bands are shown.
(c) Distribution of $z_\b = 2E_\b/\sqrt{s}$ at the parton level.
(d) Distribution of $z_\B = 2E_\B/\sqrt{s}$, i.e. the fragmentation
function, at the hadron level (only primary hadrons are considered). 
}\label{fig:higgs}
\end{figure}

We choose to study the production of $\b\bbar$ pairs at a CM energy of
130 GeV. This number is in the middle between the current lower limit
and the $\W^+\W^-$ threshold, at around the value expected for the MSSM
$\mathrm{h}^0$, but obviously the relative comparison of the sources is
only mildly energy-dependent. Five different sources are compared. The
first sample is considered as a reference and consists of gluon radiation 
in the decay $\gamma^*/\Z^* \to \q\qbar$, where q is a light u, d or s
quark. For simplicity they are assumed to be produced in equal amounts.
The other four samples consist of gluon radiation in the decay
$X \to \b\bbar$, where $X$ is a vector, axial vector, scalar or
pseudoscalar source. Clearly the $\gamma^*/\Z^*$ is a mixture,
eq.~(\ref{gamZmix}), and also $\mathrm{h}^0/\H^0/\A^0$ need not be pure
states, but the separation allows us to study the extreme range of
possibilities.

Fig.~\ref{fig:higgs}a shows the angular energy flow. The difference
between having or not having a $\gamma_5$ in the coupling is negligible,
but the radiation at large angles is larger for the spin 0 source than a
spin 1 one. The measure in Fig.~\ref{fig:higgs}b, i.e. the normalized
3-jet rate, is most sensitive and shows that the 3-jet rate is
significantly larger for a spin 0 source, especially for well separated
jets (large $y_c$), which is consistent with the larger energy flow at
large angles. The parity dependence is very small, on the other hand,
and is neglected in the following.

Considering the large spin dependence one could expect the fragmentation
function also to change considerably when going from a vector to a scalar 
source. Fortunately, as can be seen in Figs.~\ref{fig:higgs}c--d, the
changes are minor, indicating that the fragmentation function is mainly
sensitive to the bulk of gluons at smaller angles, where the sources give 
the same emission rate. The larger gluon energy flow for a scalar source
is reflected in a slightly smaller $\langle z \rangle$, with a difference 
of about 1\% both at the parton and hadron level. The small effect on the 
fragmentation function is positive from an experimentalist's point of
view, in that fragmentation functions measured at the $\Z^0$ pole can be
simply extrapolated also for possible spin 0 sources at higher energies.

\subsection{Top production and decay}
\label{sec:top}

For top production around the $\t\tbar$ threshold, real gluon emission
will be limited because of phase space, so aspects such as the form of
the QCD confinement potential, best tackled with other perturbative
techniques, will be more relevant \cite{topatthresh}. In this section we
study the process $\e^+\e^- \to \t \tbar$ at 500 GeV, where gluon emission
will start to become important, and compare the new shower routine with
the older versions.  We focus on fundamental aspects such as gluon
radiation patterns and top mass determinations. In the process we will
also study gluon radiation in the decay of the top, and differences in the
radiation patterns for the two decay channels $\t \to \b \W^+$ and
$\t \to \b \H^+$ to demonstrate the advertised process dependence of the
new shower routine.

\begin{figure}
\begin{center}
\epsfig{file=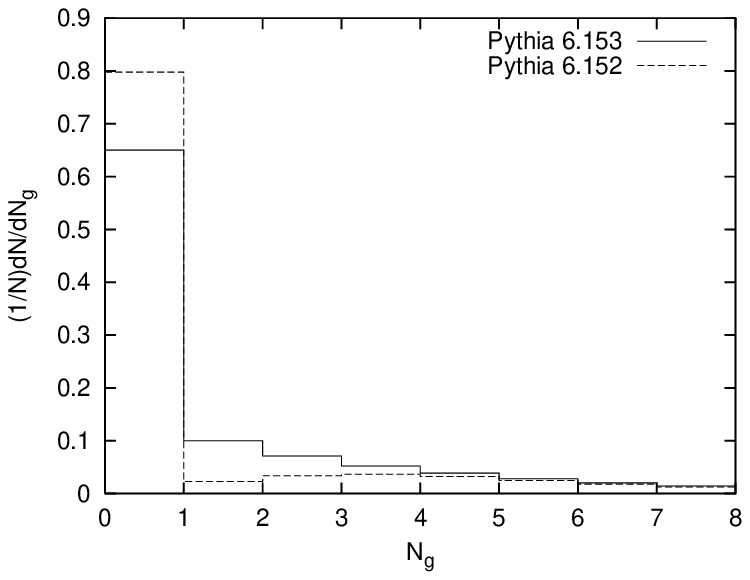}
\epsfig{file=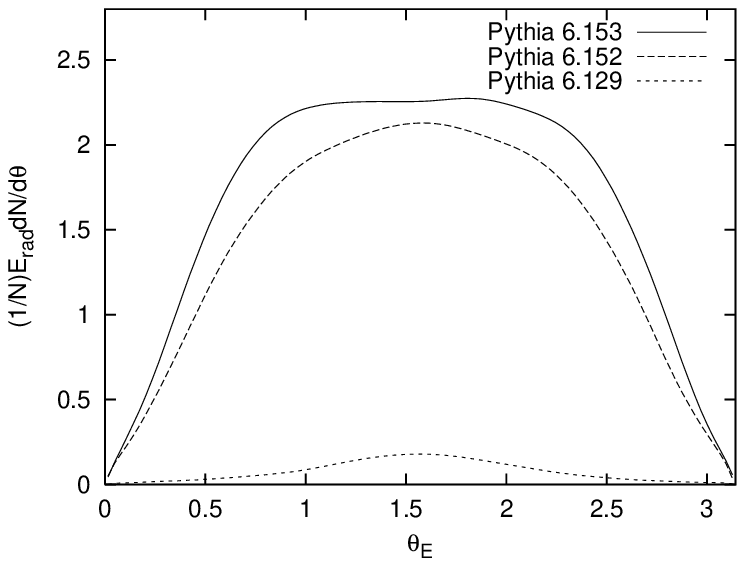}
\end{center}
\caption{Gluon multiplicity and energy flow in the $\t\tbar$ shower at
500 GeV centre-of-mass.
}\label{fig:ttbar}
\end{figure}

We first study gluon radiation in the production of the top.
The matrix element used in the correction of the shower is an energy
dependent mixture of vector and axial vector ones. High above the $\Z^0$
pole, the vector part will dominate and at 500 GeV the $\alpha$
parameter, introduced in Section~\ref{sec:calc} to parameterize the
relative mixture, is about 0.78 for $\t\tbar$ production. The dead cone
effect is expected to be very pronounced in this case, because the scaled 
mass is large, around $r=0.35$. Fig.~\ref{fig:ttbar} shows the gluon
multiplicity and the energy flow in the $\t\tbar$ shower. Most events
have no gluon emission, but the no-gluon rate has decreased in the new
shower as expected. The result of an even older version, where the dead
cone effect is severely exaggerated in all branchings, is shown as a
reference. The lesson is that the correct description of the first
emission, i.e. the difference between 6.129 and 6.152, is much more
important than that of subsequent emissions, between 6.152 and 6.153.
Once a gluon is emitted, it tends to be quite energetic and splits into
several further gluons. This is the reason for the large spread in the
multiplicity and the dip in this distribution for the old shower routine. 
The total energy flow has also increased in the new shower relative to
the older ones. Even if the energy flow seems to be large at large angles,
most energy is kept by the original top quarks at
$\theta_\mathrm{E}=0~\mathrm{and}~\pi$ respectively.

The dominant top decay is $\t \to \b \W^+$, where further gluons will be
radiated. In the old shower routine, gluon radiation off the b did not
take into account the full t--b interference structure. Furthermore,
gluon radiation in a hypothetical decay $\t \to \b \H^+$ was the same as
in the one above, apart from the difference induced by the different W
boson and Higgs masses. In order to separate off this trivial mass effect
we put $m_\H=m_\W$, well aware that this is not realistic.

\begin{figure}
\begin{center}
\epsfig{file=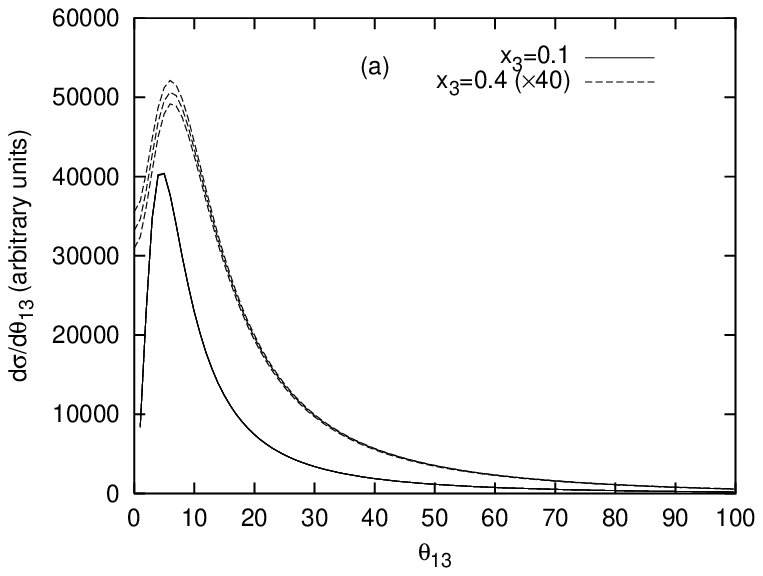}
\epsfig{file=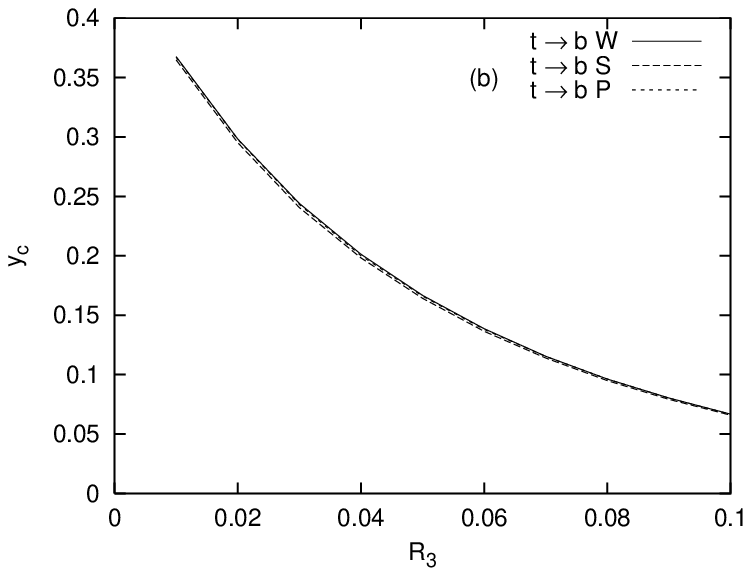}
\epsfig{file=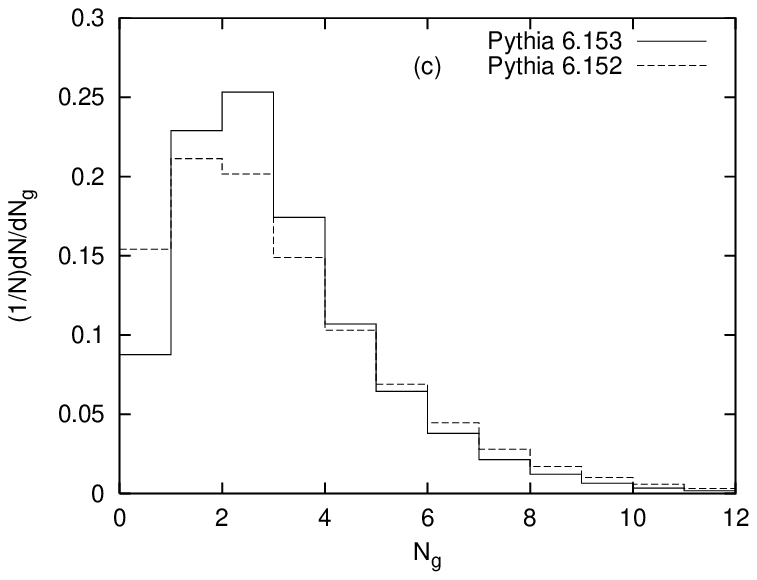}
\epsfig{file=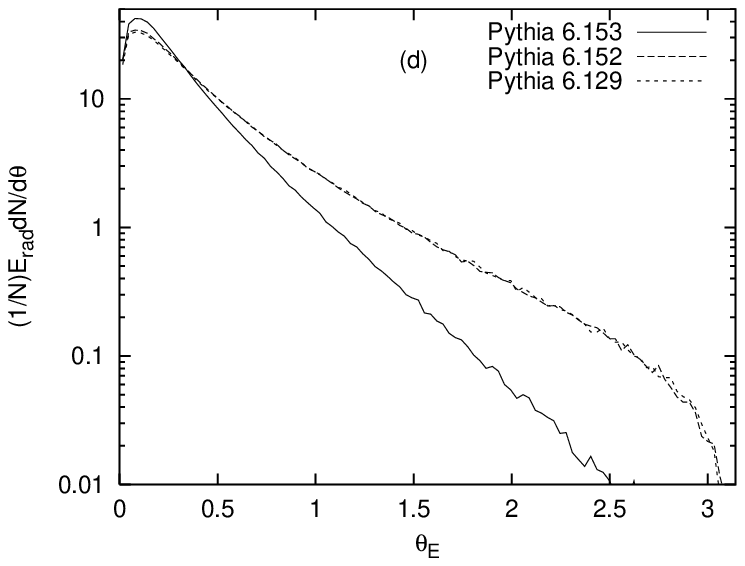}
\end{center}
\caption{(a) The gluon emission rate as a function of the emission angle
$\theta_{13}$ for top decay, $\t \to \b X$. The emission rate is given
for the two gluon energies $x_3=0.1$ (full curves) and $x_3=0.4$ (dashed).
The curves corresponding to the higher energy have been multiplied by a
factor of 40 for clarity. Three different curves are shown for each
energy: pseudoscalar Higgs, W boson and scalar Higgs (from high to low
curves). (b) The 3-jet rate in complete $\t\tbar$ events as a function of 
$y_c$ for the three different cases above, where the W/H are removed from 
the cluster search. (c--d) Gluon multiplicity and energy flow in the
shower induced by the top decay (in the rest-frame of the decaying top
with the $\W$ at $\theta_\mathrm{E}=\pi$).
}\label{fig:topdecay}
\end{figure}

The matrix element used in the decay $\t \to \b \W^+$ has the familiar
$V-A$ coupling, but the $\t \to \b \H^+$ case is less obvious. In the
MSSM the vertex factor is proportional to
$m_\t(1-\gamma_5) + m_\b\tan^2\beta(1+\gamma_5)$,
i.e. a parameter-dependent mixture of scalar ($S$) and pseudoscalar ($P$)
couplings. Naively, the $S-P$ one would be expected to dominate because
of the large difference between the top and bottom masses. However, the
favoured values of $\tan\beta \gsim 3$ could cancel this effect and give
rise to an almost pure scalar coupling. We therefore study the two
extreme cases of $S$ and $P$ separately. The difference between the
matrix elements of this type has been studied in
Tables~\ref{tab:threejetrate} and \ref{tab:threejetratezero} and found to
be non-negligible for massive daughters but to vanish in the massless
limit. For the process $\t \to \b X$, where $X$ is a $\W^+$, pure scalar
$\H^+$ or pure pseudoscalar $\H^+$, some differences are visible in the
gluon energy flow for larger gluon energies,
see Fig.~\ref{fig:topdecay}a, but the differences induced in observables
such as the jet rates are small, Fig.~\ref{fig:topdecay}b. That is, we
are closer to the massless case of Table~\ref{tab:threejetratezero},
indicating that the small mass of the radiating $\b$ is more important
than the large mass of the non-radiating $\W^+/\H^+$. In what follows,
therefore, only the decay to $\W^+$ is shown. This is the most important
case and, as it turns out, the intermediate alternative,
cf. Fig.~\ref{fig:topdecay}a.

Figs.~\ref{fig:topdecay}c--d show the gluon multiplicity and energy flow
in the decay of the top for the new and the older shower routines. The
gluon multiplicity is slightly more peaked in the new shower and the
no-gluon rate has been somewhat reduced. The angle $\theta_E$ is defined
as the angle of emitted energy to the primary bottom quark in the rest
frame of the decaying top, which is why no peak appears at large angles,
corresponding to the direction of the colour neutral W/H. In the new
correction to the shower, gluon radiation at large angles is more
severely suppressed. As a consequence, there is more energy left for
radiation at small ones, allowing the curves to cross there. This is an
effect in addition to the influence of the new matrix element correction
scheme, visible e.g. in Fig.~\ref{fig:energyflow}a. The difference
between decays to W and H are very small also here (not shown).
The result of \Py~6.129 is similar to 6.152, but the dead cone is
slightly more pronounced, as noted before.

\begin{figure}
\begin{center}
\epsfig{file=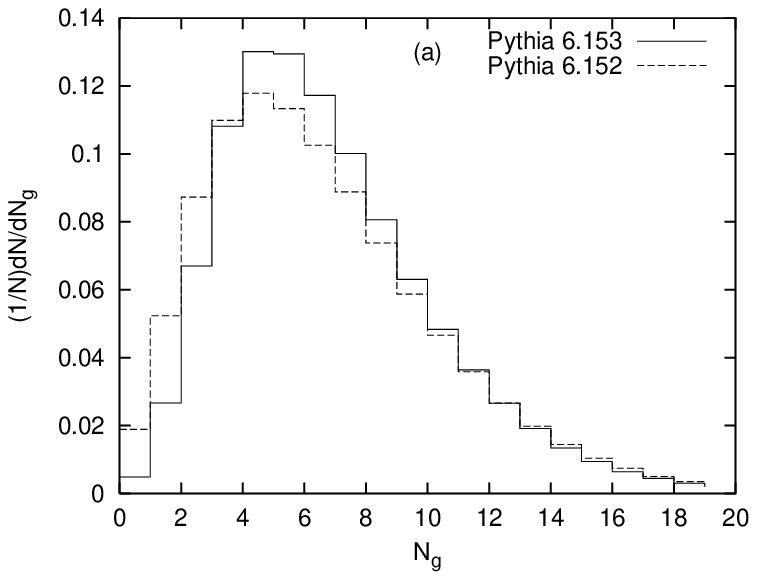}
\epsfig{file=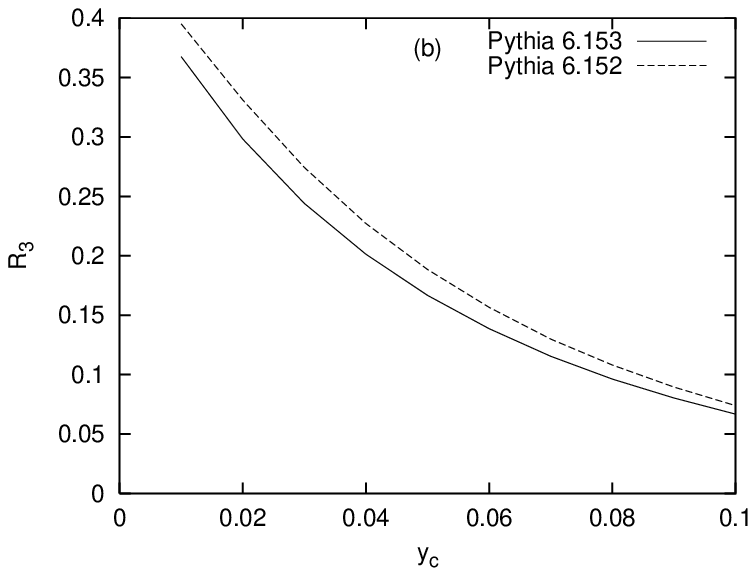}
\epsfig{file=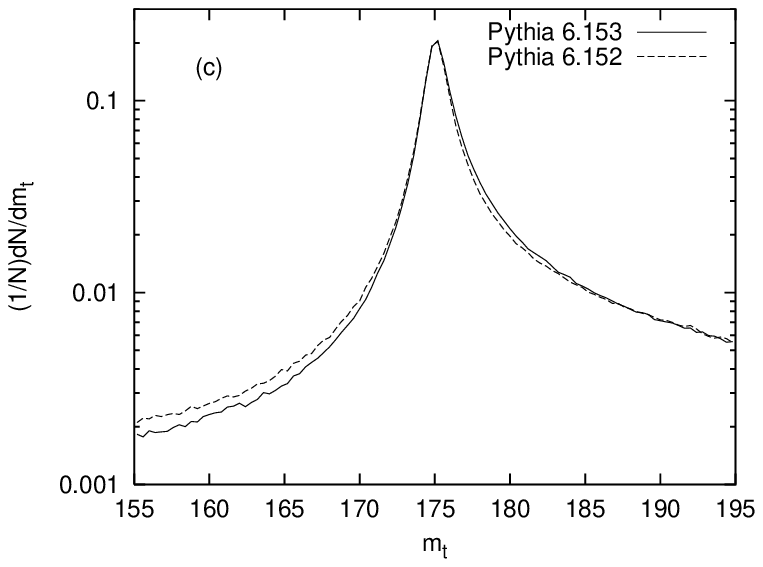}
\epsfig{file=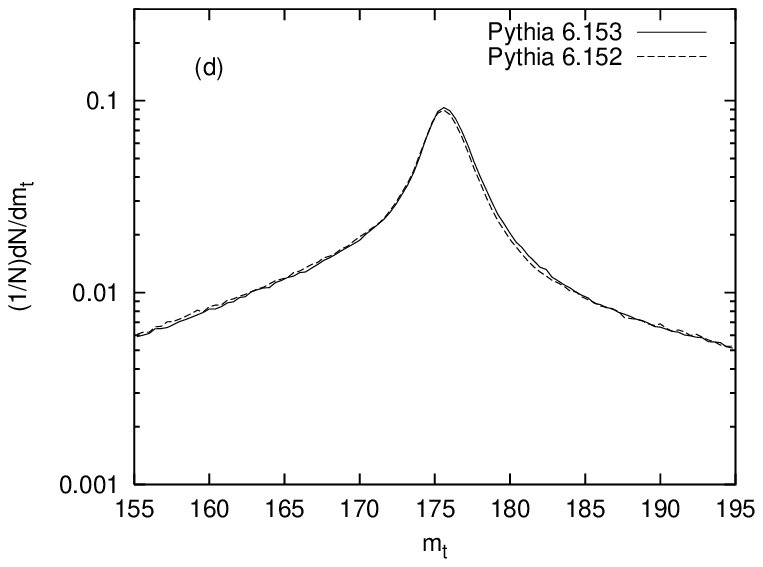}
\end{center}
\caption{$\t\tbar$ events after top decays.
(a) Total gluon multiplicity.
(b) The 3-jet rate on the parton level, with W/H removed from the
clustering. Initial-state photon radiation is included, but both models
are equally affected by this.
(c) Reconstructed top mass, parton level.
(d) Reconstructed top mass, hadron level.
}\label{fig:global}
\end{figure}

In a complete event, gluon emission from the decay and production are
added, and also possible radiation in hadronic decays of the W/H must be
considered. In principle, these showers will also interfere but, as a
realistic first approximation, we work in the zero width limit where
these interferences vanish. Considering the two stages of production and
decay together, but ignoring the decay products of the W/H,
Fig.~\ref{fig:global}a shows the total gluon multiplicity for the new and
older shower routines. Since the radiation has increased in the
production and decreased in the decay, effects tend to cancel. The net
result is a slight increase in the average gluon multiplicity and a
somewhat smaller width. The 3-jet rate, however, has decreased,
Fig.~\ref{fig:global}b, presumably because of the reduced probability to
have gluon emissions at large angles in the top decay, cf.
Fig.~\ref{fig:topdecay}d.

The total gluon multiplicity is not a very good measure of the event
properties and certainly not an observable. To give a practical example
where gluon radiation is important, we consider the reconstruction of the
top mass in a simplified scenario, assuming that the W/H can be completely
reconstructed in order only to study the effects of gluon emission in the
top production and decay. We find two jets and calculate the top mass by
considering the two possible combinations of jet+W momenta. The
combination which minimizes $(m_1-m_\t)^2+(m_2-m_\t)^2$, where
$m_t=175$~GeV, gives the reconstructed top masses $m_1$ and $m_2$. The
distribution of the reconstructed mass is given in
Fig.~\ref{fig:global}c, where jets are clustered on the parton level.
The reduced radiation level survives in this measure as a slightly
narrower peak for the new shower. The effects of fragmentation and decay
are quite large for the reconstructed mass, but some differences still
survive in the high mass wing, as seen in Fig.~\ref{fig:global}d.

\subsection{Supersymmetry production and decay}
\label{sec:susy}

In supersymmetric models, a whole new set of processes involving coloured
particles is introduced. We have calculated the most important LO matrix
elements of the MSSM, listed in Table~\ref{tab:processes}, to be used as
input to the parton shower routine. In versions of \Py~prior to 6.153,
supersymmetric particles did not shower. Because of the large masses of
these hypothetical particles, this is a good first approximation.
However, if supersymmetric particles are discovered at the LHC or a
future linear collider, detailed studies of their properties will profit
from a better understanding of QCD radiation patterns.

A complication that appears in the MSSM is significant rates of three
body decays involving one or several coloured particles in the initial or
final states. Often several interfering diagrams with intermediate
off-shell propagators contribute. An example is the decay
$\tilde{\chi}^+ \to \tilde{\chi}^0 \q'\qbar$ with either a $\W$ or squark
propagator. These processes do not fit into the present framework with
corrections to matrix elements of the type $a \to b c \g$.
As a preliminary solution, the $\q'\qbar$ pair above is assumed to be
produced from a $V-A$ source. A new kind of gluon radiation process in
the MSSM is from four-vertices such as
$\tilde{\q} \to \tilde{\q}' \W^+ \g$. These diagrams introduce no new
divergences. The interference terms only contribute to the ordinary
divergences, and the squared amplitude gives rise to a constant.
An upper estimate is again given by the new shower expression, which is
corrected by the full matrix element.

As a first example, consider top vs. stop production at a linear
\ee~collider and the decay $\t \to \b \W^+$ vs.
$\st \to \b \tilde{\chi}^+$. Again, identical masses are assumed and
consequently $m_{\st}=m_\t$ and $m_{\tilde{\chi}}=m_\W$ are used. As
usual, the $\t\tbar$ source is an energy dependent mixture of vector and
axial vector. Only a scalar coupling is possible for
$\Z^0 \to \st \stbar$, so no ambiguity here either. The top decay has the
$V-A$ coupling, while the stop one is a parameter dependent mixture of
scalar and pseudoscalar. Again we consider the two extreme cases
separately. While the stop obviously decays isotropically, the current
\Py~implementation does not include top polarization information and thus
also decays this particle isotropically. Initial-state photon radiation
has not been included in the comparison, since the different threshold
behaviours of the $\t\tbar/\st\stbar$ cross sections allow more energetic
radiation in the former process, which will affect event topologies.

\begin{figure}
\begin{center}
\epsfig{file=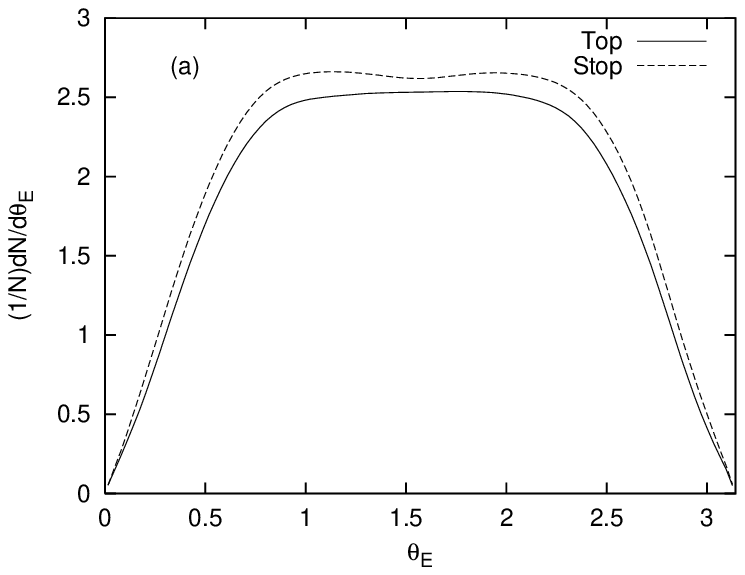}
\epsfig{file=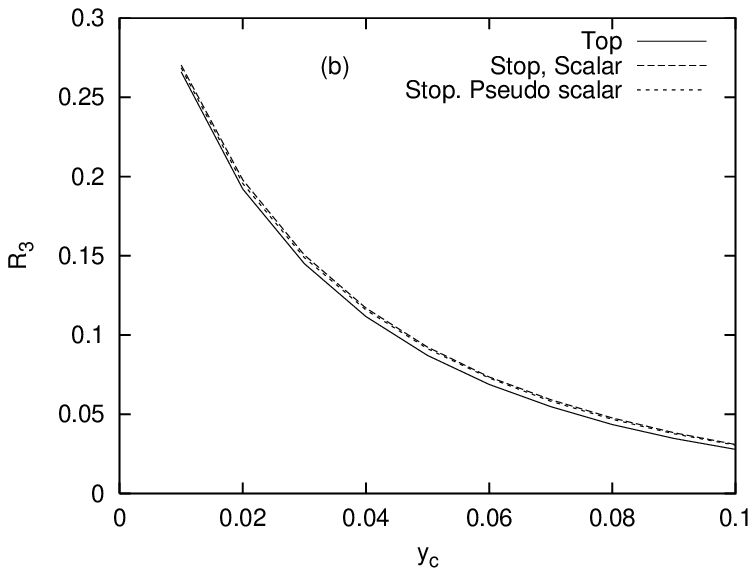}
\end{center}
\caption{Top vs. stop production and decay at a 500 GeV \ee~collider.
(a) energy flow in the $\t\tbar$ ($\st\stbar$) shower.
(b) 3-jet rate on the parton level for the production and decay taken
together, with the $\W/\tilde{\chi}$ removed from the clustering.
}\label{fig:stop}
\end{figure}

Fig.~\ref{fig:stop}a shows the energy flow for top vs. stop production.
The multiplicity is the same for both top and stop, but the energy flow
is slightly larger for stop at large angles. The difference between the
gluon radiation patterns in the top and stop decays is negligible
(not shown). As a result of the increased gluonic energy flow, the 3-jet
rate at the parton level, with the $\W/\tilde{\chi}$ removed from the
cluster analysis, is slightly larger for stop, Fig.~\ref{fig:stop}b.
We notice further that the parity dependence is negligible also in this
case. The differences between supersymmetric and standard model processes
is not large in this example, at least when mass effects are neglected,
and the ambiguity in the coupling structure is negligible. This is all
good news, but since gluon emission from supersymmetric particles is a
completely new feature, many more tests should be done.

As a last example we consider a full simulation of gluino pair production
(via $\g\g \to \tilde{\g}\tilde{\g}$) at a 14 TeV $\p\p$ collider. Since
we have not developed a full shower formalism for $2 \to 2$ processes,
as a first approximation the eikonal expression will be used for
radiation off the $\tilde{\g}\tilde{\g}$ system, including a colour
factor rescaling by $N_C/C_F=9/4$. Initial state radiation is studied
separately and multiple interactions are neglected. A scenario with
$m_{\tilde{\g}}=450~\mathrm{GeV}$, $m_{\tilde{\b}_1}=250~\mathrm{GeV}$
and $m_{\tilde{\t}_1}=200~\mathrm{GeV}$ is used. The dominant gluino
decays are then $\tilde{\g} \to \tilde{\b}_1 \bbar~(\tilde{\t}_1 \tbar)$
and the squarks decay by processes of the type $\sq \to \tilde{\chi} \b$.
If the $\tilde{\chi}$ is a neutralino, which here is the lightest
supersymmetric particle, it will not decay further. A chargino decays to
a neutralino plus either leptons or quarks in approximately equal amounts.
This MSSM scenario is implemented in the \Py~event
generator~\cite{Pythia} and the details are described in the
{\sc Spythia} manual~\cite{Spythia}.

\begin{figure}
\begin{center}
\epsfig{file=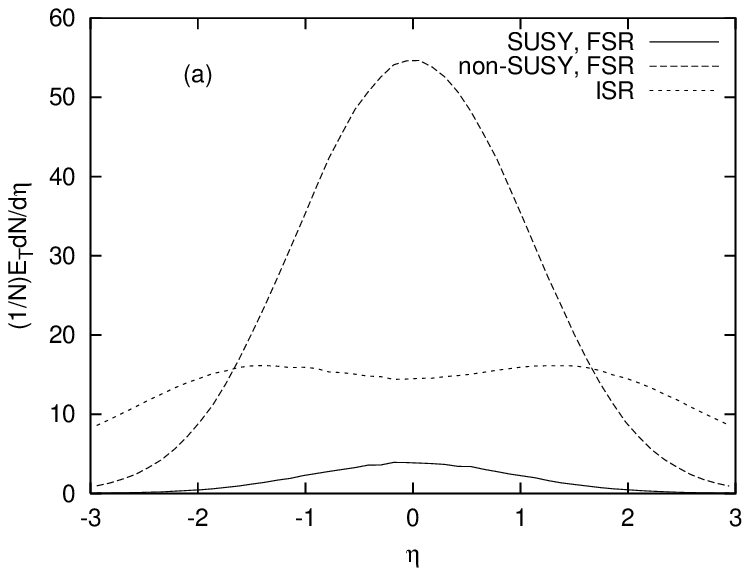}
\epsfig{file=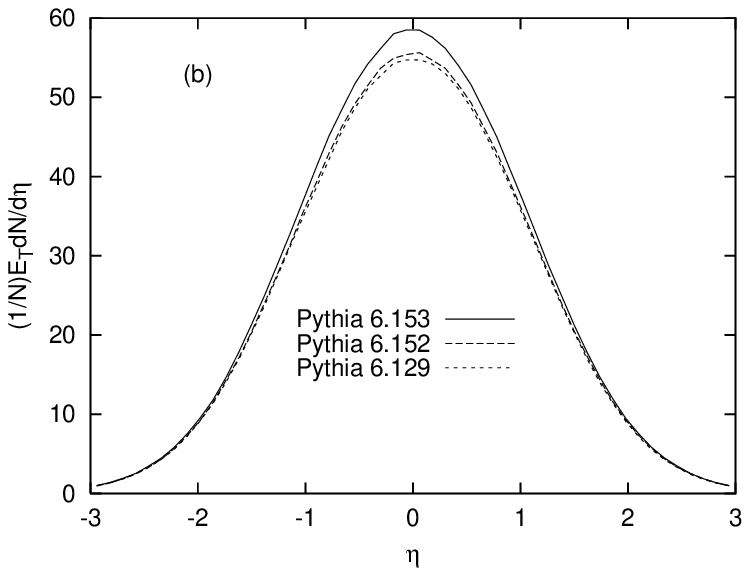}
\epsfig{file=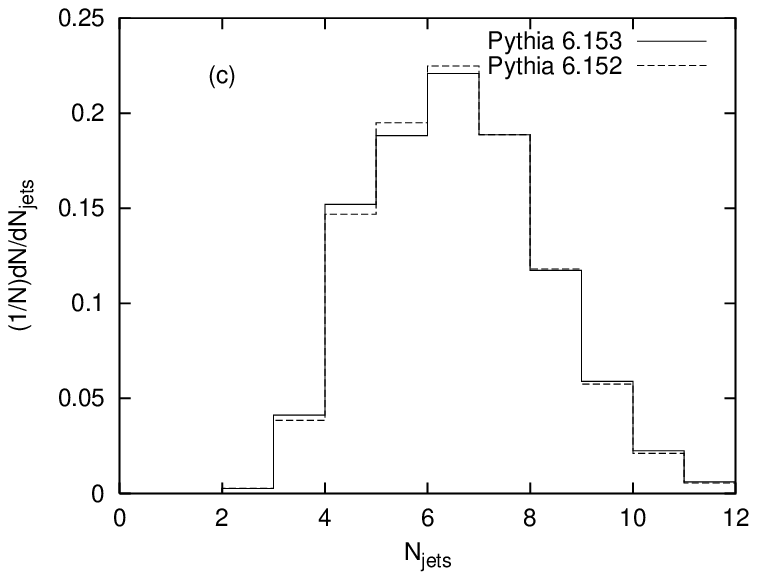}
\epsfig{file=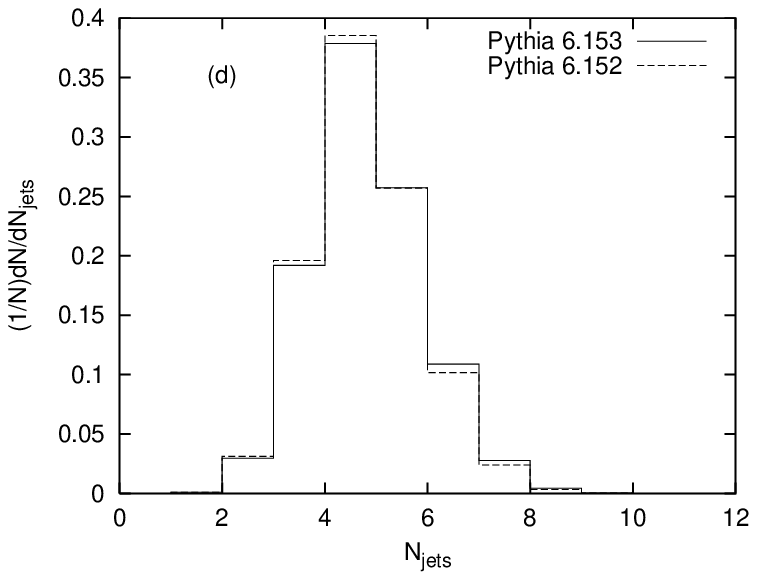}
\end{center}
\caption{Gluino production at a 14 TeV $\p\p$ collider.
(a--b) Transverse gluon energy flow in pseudorapidity,
$\eta = \frac{1}{2}\ln((p+p_z)/(p-p_z))$.
(c--d) Jet multiplicity for $R=0.75$, $E_{\perp,min}=10$
and $40~\mathrm{GeV}$ respectively.
}\label{fig:gluino}
\end{figure}

To assess the importance of gluon radiation off supersymmetric particles
in a hadron-hadron collision, we compare the transverse gluon energy
flow from SUSY and non-SUSY particles in the final state radiation (FSR),
Fig.~\ref{fig:gluino}a. The additional gluon radiation from coloured SUSY
particles is thus seen to be small compared to the ordinary gluon
radiation from quarks. In this paper we have only considered QCD final
state radiation in \ee~annihilation. In a hadron collider environment,
gluons can be radiated also from the incoming partons (initial state
radiation, ISR). This contribution is almost as large as the FSR one,
Fig.~\ref{fig:gluino}a, but the rapidity distribution is different.

In Fig.~\ref{fig:gluino}b the three \Py~versions considered in this
paper are compared. Only the FSR component is shown, and the additional
radiation from SUSY particles in the latest version gives rise to an
increased total transverse gluon energy flow. This increase is, however,
slightly compensated by the decrease in gluon radiation in the top and
squark decays, cf. Section~\ref{sec:top}. In Figs.~\ref{fig:gluino}c--d
the result of a simple cluster search is shown. Particles with a summed
transverse energy, $\sum E_\perp>E_{\perp,min}$ inside a
$\sqrt{(\Delta\eta)^2+(\Delta\varphi)^2}<R$ cone are joined in a cluster.
For large $E_{\perp,min}$, the jet multiplicity is slightly increased
when gluon radiation from supersymmetric particles is included. This
could be caused by energetic gluons radiated off the gluinos. For small
$E_{\perp,min}$, on the other hand, smaller structures are probed
and the radiation off b and lighter quarks is most important. In summary, 
we conclude that gluon radiation off supersymmetric particles at a hadron 
collider is small compared to the other sources, and of importance mainly 
in high-precision studies.

\section{Gluon splitting to heavy quarks}

Data at LEP1 show a larger rate of secondary charm and bottom 
production than predicted in most shower descriptions 
\cite{QCDWG, Mangano}, or in analytical studies \cite{Mikebc}. 
We therefore comment on a few of the issues in the Monte Carlo 
simulation and how a relaxation of some demands would affect rates.

\subsection{Strong coupling argument and kinematics}

The default behaviour in \textsc{Pythia} is to let $\alphas$ 
have $\pT^2$ as argument. Actually, the exact kinematics 
has not yet been reconstructed when $\alphas$ is invoked, so
the approximate expression $\pT^2 \approx z(1-z)m_a^2$ is used, 
see discussion at eq.~(\ref{pTbranching}). Since $\alphas$ blows up 
when its argument approaches $\Lambda_{\mathrm{QCD}}$, this 
translates into a requirement on $\pT^2$ or on $z$ and $m_a$, 
restricting allowed emissions to $\pT > Q_0/2$, where $Q_0$ is 
the shower cut-off scale. Also when full kinematics is reconstructed,
this is reflected in a suppression of branchings with small $\pT$. 
Therefore, in $\g\to\q\qbar$ branchings analyzed in the $\g$ rest 
frame, the quarks do not come out with the $1+\cos^2\theta$ angular 
distribution (with respect to the direction of motion of the gluon)
one might expect away from threshold, or a somewhat 
more isotropic one closer to threshold, but are rather peaked at 
$90^{\circ}$ and dying out at $0^{\circ}$ and $180^{\circ}$.

For $\g\to \q\qbar$ branchings, the soft-gluon results that lead to 
the choice of $\pT^2$ as scale \cite{pTscale} are no longer
compelling, however. One could instead use some other
scale that does not depend on $z$ but only on $m_a = m_{\g}$,
i.e the off-shellness of the branching gluon, and remove the
$\pT$ cut. A reasonable choice, even if not unique, is to use 
$m_{\g}^2/4$, where the factor $1/4$ ensures continuity with 
$\pT^2$ for $z=1/2$. This possibility has been added as a new 
option.

\begin{table}[t]
\begin{center}
\begin{tabular}{|c|c|c|c|c|@{\protect\rule[-2mm]{0mm}{7mm}}}
\hline
$\alphas$ &  coherence & $\g \to \u\ubar + \d\dbar + \s\sbar$ (\%) & 
$\g \to \c\cbar$ (\%) & $\g \to \b\bbar$ (\%)\\
\hline
$\alphas(\pT^2)$      & full         &   14.3   &    1.26   &    0.15 \\
$\alphas(\pT^2)$      & intermediate &   14.8   &    1.27   &    0.16 \\
$\alphas(\pT^2)$      & reduced      &   21.1   &    1.92   &    0.26 \\
$\alphas(\pT^2)$      & none         &   38.8   &    3.06   &    0.31 \\
$\alphas(m_{\g}^2/4)$ & full         &   12.9   &    1.15   &    0.15 \\
$\alphas(m_{\g}^2/4)$ & intermediate &   13.3   &    1.17   &    0.15 \\
$\alphas(m_{\g}^2/4)$ & reduced      &   20.0   &    1.78   &    0.28 \\
$\alphas(m_{\g}^2/4)$ & none         &   43.3   &    3.47   &    0.46 \\
\hline
\end{tabular}
\caption{The rate of gluon splitting to $\q\qbar$ pairs, in $\Z^0$ 
decay at 91.2 GeV with the normal primary flavour mixture.}
\label{tab:seccb}
\end{center}
\end{table}

Actually, the change of $\alphas$ argument in itself
leads to a reduced $\g\to \q\qbar$ splitting rate, while 
the removal of the $\pT > Q_0/2$ requirement increases it. 
The net result is a decrease by about 10\%, Table~\ref{tab:seccb}. 
The topologies of the events are changed somewhat, so rates 
within experimental cuts could be more affected. However, the 
changes are not as big as might have been expected --- see the 
following.

\subsection{Coherence}

In the above subsection, it appears as if the $1+\cos^2\theta$
distribution would be recovered in the new $\alphas(m_\g^2/4)$
option. However, this neglects the coherence condition, which is 
imposed as a requirement in the shower that successive opening 
angles in branchings become smaller. Such a condition actually 
disfavours branchings with $z$ close to 0 or 1, since the opening 
angle becomes large in this limit, eq.~(\ref{angle}). It should be 
noted that the opening angle discussed here is not the true one, 
but the one based on approximate kinematics, including neglect of
masses. More generally, the coherence formalism is not really
developed with this kind of configurations in mind, especially not
with a heavy quark pair close to threshold.

As a means to exploring consequences, two new coherence level 
options have thus been introduced. In the first, the $\pT^2$ of a
$\g\to \q\qbar$ branching is reduced by the correct mass-dependent 
factor, $1-4m_{\q}^2/m_{\g}^2$, while the massless approximation is 
kept for the longitudinal momentum. This is fully within the 
uncertainty of the game, and no less reasonable than the default.
In the second, no angular ordering at all is imposed on 
$\g\to \q\qbar$ branchings. This is certainly an extreme scenario, 
and should be viewed with caution. However, it is still 
interesting to see what it leads to.

In Section~\ref{sec:additional} another ``intermediate'' coherence variation of
the default ``full'' one was introduced, affecting the rate of
gluon emission off the primary quarks but not the subsequent gluon
cascades. This variation has negligible consequences for the
secondary quark rate at LEP energies and is not considered further.

It turns out that the decay angle distribution of the gluon is 
much more distorted by the coherence than by the $\alphas$ 
and kinematics considerations described earlier. Both 
modifications are required if one would like to have a 
$1+\cos^2\theta$ shape, however. Also other distributions, like 
gluon mass and energy, are affected by the choice of options.

The most dramatic effect appears in the total gluon branching
rate, however, Table~\ref{tab:seccb}. Already the reduced  angular 
ordering requirement can boost the $\g\to \b\bbar$ rate by almost a 
factor of two. The effects are even bigger without any angular 
ordering constraints at all. It is difficult to know what to make of 
these big effects. Experimental information on the angular 
distribution of secondary $\c\cbar / \b\bbar$ pairs might help 
understand what is going on, but probably that is not possible 
experimentally. Anyway, the measured (but still uncertain) values
are fully bracketed by the range of the models, indicating that
there need not be a conflict between theory and experiment.

\section{Summary and Outlook}

We have in this article studied QCD radiation off heavy particles,
both based on the calculation of a wide set of first-order matrix
elements and by the usage of these matrix elements as input to 
a parton-shower description of multiple-gluon emission.

The matrix-element calculations provide at least two important 
insights. One is the significant spin dependence on the
rate of gluon emission, given identical kinematical
conditions. Some such effects could be expected simply from the
different shape of the splitting kernel of a quark and a squark,
say, but equally important is the spin of the decaying particle,
and the parity of the process (combining that of mother and
daughters). The best illustration is the difference between a
vector and a scalar colour singlet source each decaying to a pair
of squarks, where the gluon emission rate differs by up to a factor of 1.8
in our limited study, and is likely to reach even higher levels in 
other observables. The $\gamma_5$ dependence disappears in the  
massless limit, but the bulk of other effects remain also there,
so this is not only an issue for the production of heavy flavours.
This process dependence would not be caught by the traditional 
parton-shower philosophy, where the universal aspects of gluon 
emission are emphasized.    

The other insight is that the `dead cone' concept is one that only
applies universally for soft gluon emission. Only one process provides 
an exact zero in the collinear emission rate while, for reasonably 
hard gluons, some processes show no dip at all at small angles. And, 
while soft gluons may be ideologically interesting, the experimentally 
observable event shapes are often more crucially dependent on some 
intermediate range of gluon energies. Again, therefore, the message is 
that a universal shower description may be misleading.  

Fortunately, it is not impossible to combine the shower picture
with a process dependence of the kind noted above. We have in this
article developed one approach to the problem, wherein matrix element
information is provided for all shower branchings of the primary
particles. This combines a process dependent resummation of gluon 
emission off these particles with the traditional strengths of the
shower formalism, such as providing exclusive final states with exact 
energy--momentum conservation.

To illustrate this formalism, a few physics examples have been studied.
For bottom production at the $\Z^0$ pole, detailed comparisons with
data are possible. We find the new shower routine to be in good agreement
with current data if the uncertainties in both model and data are taken
into account. A significant dependence on coherence options were found both for
jet rates and gluon splitting rates in the shower. It has further been possible to
rule out an older version of the shower routine where mass effects were not
correctly accounted for.

Differences in jet rates were found for different coupling
structures in Higgs decay to bottom, whereas the influence on the fragmentation
function is minor. Both the production and decay of the top quark is
dependent on the matrix element correction, influencing e.g. the mass reconstruction
of the top quark. Again the jet rate was found to be most sensitive.

Gluon radiation off supersymmetric particles, squarks and gluinos, has
been introduced, but the effect of this additional radiation is small
in high energy processes, especially at hadron colliders, because of
the background from initial state radiation and showers off the standard
model decay products. Differences between top and stop events are small
if equal masses are assumed, but many more comparisons between standard
model and supersymmetric processes could be envisaged.

From this limited study we conclude that gluon emission off b
and lighter quarks dominate the picture. If this component is
modelled well, the rest is less important, but still of ideological
interest. Since the scaled bottom mass $r=m_\b/m_X$, in the decay of
a resonance $X$, is small in most cases, we see little if any
parity dependence. In $\Z^0$ decays this effect is less than 1\%,
but it could be visible in a high-precision study. Also
for the spin dependence, effects are small in most cases,
although the $\H^0 \to \b\bbar$ example shows they need not
be quite as small. It thus turns out that the new process dependence
will mainly be important when high precision is strived at.

We note that the current study is not the end of the story. While we
have sampled a fair selection of colour and spin structures, more
are likely to turn up even in the simple context of two-body decays
considered here. The MSSM also allows a significant rate for 
three-body decays, normally as a sequence of two consecutive two-body
ones with an intermediate off-shell propagator, and with the possibility
of interference between several such intermediate states. The real
challenge, however, may well be provided by the more complex production
processes at hadron colliders, where the concept of a sequence of 
$s$-channel processes need no longer be valid, e.g. squark production 
$\g\g\to\sq\sqbar$ with a $t$-channel squark propagator. The large 
rate of initial-state radiation also implies that the initial--final
state interference terms may be more important than the final-state 
radiation off the heavy particles themselves. Much work therefore lies 
ahead, if one desires a good description of QCD effects in many 
processes of interest at the LHC.

\end{document}